\documentclass[a4paper,fleqn,usenatbib]{mnras2}
\usepackage{newtxtext,newtxmath}
\usepackage[T1]{fontenc}
\usepackage{ae,aecompl}
\usepackage{graphicx}
\usepackage{appendix}

\usepackage{ulem}
\usepackage{amsmath}
\usepackage{amssymb}
\usepackage{natbib}
\usepackage{mathtools}
\usepackage{threeparttable}
\usepackage{booktabs}
\usepackage{verbatim}
\usepackage{hyperref}
\usepackage{color}
\usepackage{bm}
\defcitealias{2023MNRAS.526..932G}{Paper~I}

\title[EC-SNe and the DNS systems]{Electron-capture supernovae in NS+He star systems and the double neutron star systems}

\author[Y. Guo et al.]{
	Yun-Lang Guo,$^{\rm 1,2}$\thanks{E-mail:yunlang@nju.edu.cn}
	Bo Wang,$^{\rm 3,4}$\thanks{E-mail:wangbo@ynao.ac.cn}
	Wen-Cong Chen,$^{\rm 5}$
	Xiang-Dong Li,$^{\rm 1,2}$\thanks{E-mail:lixd@nju.edu.cn}
	Hong-Wei Ge,$^{\rm 3,4}$\newauthor
    \,\,Long Jiang$^{\rm 5}$ and
	Zhan-Wen Han$^{\rm 3,4}$
	\\
	$^{1}$School of Astronomy and Space Science, Nanjing University, Nanjing 210023, China\\
	$^{2}$Key Laboratory of Modern Astronomy and Astrophysics, Nanjing University, Ministry of Education, Nanjing 210023, China\\
	$^{3}$Yunnan Observatories, Chinese Academy of Sciences, Kunming 650216, China\\
	$^{4}$International Centre of Supernovae, Yunnan Key Laboratory, Kunming 650216, China\\
	$^{5}$School of Science, Qingdao University of Technology, Qingdao 266525, China\\
}

\date{Accepted XXX. Received YYY; in original form ZZZ}

\pubyear{2022}

\begin{document}
\label{firstpage}
\pagerange{\pageref{firstpage}--\pageref{lastpage}}
\maketitle

\begin{abstract}
Electron-capture supernovae (EC-SNe) provide an
alternative channel for producing neutron stars (NSs).
They play an important role
in the formation of double NS (DNS) systems and the chemical evolution of galaxies,
and contribute to the NS mass distribution in observations.
It is generally believed that EC-SNe originate from $e$-captures
on $\rm^{24}Mg$ and $\rm^{20}Ne$
in the massive degenerate oxygen-neon (ONe) cores with masses close to the Chandrasekhar limit ($M_{\rm Ch}$).
However,
the origin of EC-SNe is still uncertain.
In this paper,
we systematically studied the EC-SNe in NS+He star systems
by considering the explosive oxygen burning that may occur in the near-$M_{\rm Ch}$ ONe core.
We provided the initial parameter spaces for producing EC-SNe
in the initial orbital period $-$ initial He star mass
(log$P_{\rm orb}^{\rm i}-M_{\rm He}^{\rm i}$) diagram,
and found that both $M_{\rm He}^{\rm i}$ and minimum $P_{\rm orb}^{\rm i}$ for EC-SNe increase with metallicity.
Then, by considering NS kicks added to the newborn NS,
we investigated the properties of the formed DNS systems after the He star companions collapse into NSs,
such as the orbital periods, eccentricities and spin periods of recycle pulsars ($P_{\rm spin}$), etc.
	The results show that most of the observed DNS systems
	can be produced by NS kicks of $\lesssim50\rm\,km\,s^{-1}$.
In addition,
we found that NSs could accrete more material
if the residual H envelope on the He star companions is considered,
which can form
the mildly recycled pulsars ($P_{\rm spin}\sim20$\,ms) in DNS systems.

\end{abstract}


\section{Introduction}
Electron-capture supernovae (EC-SNe) are induced by the $e$-capture on
$^{24}\rm Mg$ and $^{20}\rm Ne$ in the strongly degenerate oxygen-neon (ONe) cores
with masses close to the Chandrasekhar limit ($M_{\rm Ch}$; \citealt{1987ApJ...322..206N}).
It is generally suggested that the cores will eventually
collapse into neutron stars (NSs) after undergoing EC-SNe
\citep[e.g.][]{1984ApJ...277..791N, 2007A&A...476..893S, 2013ApJ...772..150J, 2015MNRAS.446.2599D, 2015MNRAS.451.2123T, 2023pbse.book.....T}.
Because of the relatively steep density gradient in the outer cores,
EC-SNe are expected to experience a prompt or fast explosion
\citep[see][]{2004ApJ...612.1044P}.
Thus they may have
relatively low explosion energies
\citep[$\rm \sim 10^{50}\,erg$,][]{2006A&A...450..345K},
small amounts of $^{56}\rm Ni$
\citep[$\rm \sim 0.002-0.015\,M_\odot$,][]{2006A&A...450..345K,2009ApJ...695..208W},
and low-velocity kicks
\citep[$\rm \lesssim 100\,\rm km\,s^{-1}$,][]{2018ApJ...865...61G,2020MNRAS.496.2039S}.
In view of this,
EC-SNe can explain a class of high-mass X-ray binaries
with long orbital periods and low eccentricities,
as well as the retention for NSs in globular clusters
\citep[e.g.][]{2002ApJ...573..283P,2004ApJ...612.1044P,2004ESASP.552..185V,2010NewAR..54..140V}.
Meanwhile,
EC-SNe can be used to explain some low luminosity SNe,
e.g. SNe 1997D \citep[][]{1998ApJ...498L.129T},
SN 2005cs \citep[][]{2006MNRAS.370.1752P, 2009MNRAS.394.2266P},
SN 2008S \citep[][]{2009MNRAS.398.1041B}
and SN 2018zd \citep[][]{2021NatAs...5..903H}.
They are also potential sources for the $r$-process,
which can synthesize heavy elements and contribute to the chemical
evolution of galaxies \citep[][]{2007ApJ...667L.159N}.
In addition,
\citet{2010ApJ...719..722S} studied the masses of 14 well-measured NSs,
and found that the mass distribution of NSs may be bimodal,
in which the lower mass peak ($\sim 1.25\,\rm M_\odot$)
may arise from EC-SNe.

Double neutron star (DNS) systems are important potential gravitational wave sources,
which can be formed in NS+He star systems after the He star companions
undergo EC-SNe or iron core collapse supernovae
\citep[Fe CC-SNe; e.g.][]{2002MNRAS.331.1027D, 2003MNRAS.344..629D, 2003ApJ...592..475I,  2007AIPC..924..598V, 2015ApJ...801...32A, PhysRevLett.119.161101, 2015MNRAS.451.2123T,2017ApJ...846..170T, 2018MNRAS.481.1908K, 2018MNRAS.481.4009V, 2018IJMPD..2742005H, 2021ApJ...920L..36J, 2023arXiv231202269B}.
Since the discovery of the first DNS system decades ago
\citep[PSR $\rm B1913+16$;][]{1975ApJ...195L..51H},
about 24 sources have been detected so far\footnote{ATNF Pulsar Catalogue, \href{http://www.atnf.csiro.au/research/pulsar/psrcat}{http://www.atnf.csiro.au/research/pulsar/psrcat}
\citep[version 1.71, November 2023;][]{2005AJ....129.1993M}}.
The orbital periods of the observed DNS systems span a wide range from $0.1$\,d to $45.0$\,d,
the orbital eccentricities range from $0.064$ to $0.828$,
and the component masses mainly range from $1.1-1.5\rm\,M_\odot$.
The observed features of DNSs are important for exploring their origin
\citep[e.g.][]{2017ApJ...846..170T, 2018ApJ...867..124S, 2019ApJ...880L...8A}.
It seems that most DNS systems are produced by small NS kick velocities,
indicating that the EC-SN channel may provide an important contribution to the formation of DNS binaries.
However, there are still some uncertainties for the origin of EC-SNe,
although they play a vital role in
the formation of DNS binaries and the NS mass distribution in observations
\citep[e.g.][]{2015MNRAS.451.2123T, Jones2016, 2022A&A...668A.106C}.
 
Many works have been devoted to the study of the initial mass range for producing EC-SNe,
which can be used in population synthesis studies
\citep[e.g.][]{1984ApJ...277..791N,2007PhDT.......212P,2015MNRAS.446.2599D,2015MNRAS.451.2123T,2020RAA....20..161H}.
It is generally believed that the initial masses of single main-sequence stars
for producing EC-SNe are in the range of $\sim 8-10\rm\,M_\odot$
for solar metallicity,
and this mass range is metallicity dependent
\citep[see e.g.][]{2017PASA...34...56D}.
In the classic picture,
the lower limit of metal core mass for producing EC-SNe is $\sim 1.37\,\rm M_\odot$
\citep[e.g.][]{1984ApJ...277..791N,2013ApJ...771...28T},
and the upper limit of metal core mass is $\sim 1.43\,\rm M_\odot$
\citep[][]{2015MNRAS.451.2123T}.
For the metal cores with masses less than $\sim 1.37\,\rm M_\odot$,
stars will evolve into ONe white dwarfs (WDs).
For the metal cores with masses larger than $\sim 1.43\,\rm M_\odot$, the stars will collapse into NSs through Fe CC-SNe.
For the metal cores with masses of $1.37-1.43\,\rm M_\odot$,
the oxygen deflagration is initiated in the center 
due to the heating of $e$-capture on $^{20}\rm Ne$.
Subsequently,
as the central temperature rises steeply,
nuclear statistical equilibrium (NSE) will be achieved
when temperature exceeds $5\times10^9$\,K.
The $e$-capture reactions in NSE region accelerate core contraction,
resulting in the formation of NSs eventually
if the $e$-capture rate exceeds the nuclear burning rate.

However,
\citet[][hereafter \citetalias{2023MNRAS.526..932G}]{2023MNRAS.526..932G} recently found that explosive oxygen burning may be triggered
after He stars develop ONe cores with masses close to $M_{\rm Ch}$,
and then the He stars may explode as Type Ia supernovae (SNe Ia)
instead of collapsing into NSs
\citep[see also][]{Antoniadis2020A&A, 2022A&A...668A.106C}.
It is interesting to see
how these new results affect the evolution of binary NSs resulting from EC-SNe.
Accordingly,
we will explore the formation of the EC-SNe in NS+He star binaries,
and the properties of DNSs originating from the EC-SN channel.
The basic method and assumptions for the binary evolution simulations
are introduced in Section 2.
The simulation results for EC-SNe in NS+He star binaries are presented in Section 3.
We show the investigation of the properties of DNSs produced by EC-SNe in Section 4.
Finally, we give discussions in Section 5 and a summary in Section 6.
\section{Numerical methods and assumptions}
We performed detailed binary evolution calculations of NS+He star systems
by using the stellar evolution code Modules for Experiments in Stellar Astrophysics
(MESA, version 10398; see \citealt{2018ApJS..234...34P}),
in which the initial NS mass ($M_{\rm NS}^{\rm i}$) is set to be $1.35\rm\,M_\odot$.
Instead of solving the stellar structure equations of NSs,
we assume that NSs to be point masses.
In the present work, we constructed zero-age main-sequence (ZAMS) He star models
based on suite case create\_zams.
We mainly explored the parameter space for producing EC-SNe,
where the initial He star mass ($M_{\rm He}^{\rm i}$) ranges from $\sim 2.4\rm\,M_\odot$ to $3.0\rm\,M_\odot$
with a resolution of $0.01\rm\,M_\odot$,
and the initial orbital period ($P_{\rm orb}^{\rm i}$) ranges from $\sim 0.07$\,d to $10$\,d
with $\Delta$log\,$P_{\rm orb}^{\rm i}=0.5$.

During the binary evolution,
the NS starts to accrete material from He star companion once the companion fills its Roche lobe.
The scheme proposed by \citet{kolb1990A&A} is adpoted to compute the mass-transfer rate
\citep[see also the Appendix in][]{2010ApJ...717..724G},
and the fraction of mass lost from the vicinity of the NS is set to be $0.5$.
Accordingly,
the mass increase rate of NSs is
$\dot M_{\rm NS}$ $=$ min($0.5\times\dot M_{\rm tran}, \dot M_{\rm Edd}$),
in which $\dot M_{\rm Edd}=3\times10^{-8}\rm\,M_\odot\,\rm yr^{-1}$
is the Eddington accretion rate
and $\dot M_{\rm tran}$ is the mass-transfer rate
\citep[e.g.][]{2002ApJ...565.1107P, 2020ApJ...900L...8C, 2021MNRAS.506.4654W}.
In this work,
the simulated mass-transfer rates are often $3-4$ orders of magnitude
larger than $\dot M_{\rm Edd}$ and therefore limited by $\dot M_{\rm Edd}$
\citepalias[e.g. \citealt{2015MNRAS.451.2123T};][]{2023MNRAS.526..932G}.
We assume that
the excess material ($\dot M_{\rm tran}-\dot M_{\rm NS}$) is ejected from the vicinity of the NS,
carrying away the specific orbital angular momentum of the accreting NS
\citep[e.g.][]{2002MNRAS.331.1027D, chen2011A&A, 2021MNRAS.506.4654W, 2022MNRAS.515.2725G}.

During the evolution of He star companions,
we adopted the `Dutch' prescription with a scaling factor of $1.0$
as the stellar wind mass-loss mechanism
\citep[e.g.][]{2009A&A...497..255G, Antoniadis2020A&A}.
We used Type 2 Rosseland mean opacity tables provided by \citet[][]{1996ApJ...464..943I},
which can be applied to the enhanced carbon-oxygen caused by He-burning.
The nuclear reaction network is coupled by $43$ isotopes from $\rm ^1H-^{58}Ni$,
including NeNa and MgAl cycles, Urca processes of
$^{23}\rm Na\rightleftharpoons$ $^{23}\rm Ne$,
$^{23}\rm Ne\rightleftharpoons$ $^{23}\rm F$,
$^{25}\rm Mg\rightleftharpoons$ $^{25}\rm Na$ and $^{25}\rm Na\rightleftharpoons$ $^{25}\rm Ne$,
and the $e$-capture chains,
i.e. $^{24}\rm Mg$$(e^-,\nu_e)^{24}\rm Na$$(e^-,\nu_e)^{24}\rm Ne$ and
$^{20}\rm Ne$$(e^-,\nu_e)^{20}\rm F$$(e^-,\nu_e)^{20}\rm O$.
We used the weak interaction rates from \citet{2016ApJ...817..163S}.
In addition,
we adopted the HELM and PC equations-of-state
\citep[e.g.][]{2000ApJS..126..501T, 2010CoPP...50...82P, 2017MNRAS.472.3390S}.

In our simulations,
we employed the Ledoux criterion and set the mixing-length parameter to be $2.0$.
We set the overshooting parameter ($f_{\rm ov}$) to be $0.014$
\citep[][]{2013ApJ...772..150J,Antoniadis2020A&A},
and considered semi-convection \citep{1983A&A...126..207L}
and thermohaline mixing \citep{1980A&A....91..175K}\footnote{The MESA inlists for this work are publicly available at \href{https://zenodo.org/record/7655310\#.Y\_LaNS-KGx9}{https://zenodo.org/record/7655310\#.Y\_LaNS-KGx9}.}.
We stop the code if the $e$-capture on $\rm^{20}Ne$ or
the explosive oxygen burning occurs.
For the models that can undergo Fe CC-SNe,
the code is stopped after the formation of Si core.

\begin{figure}
	\centering\includegraphics[width=\columnwidth*3/3]{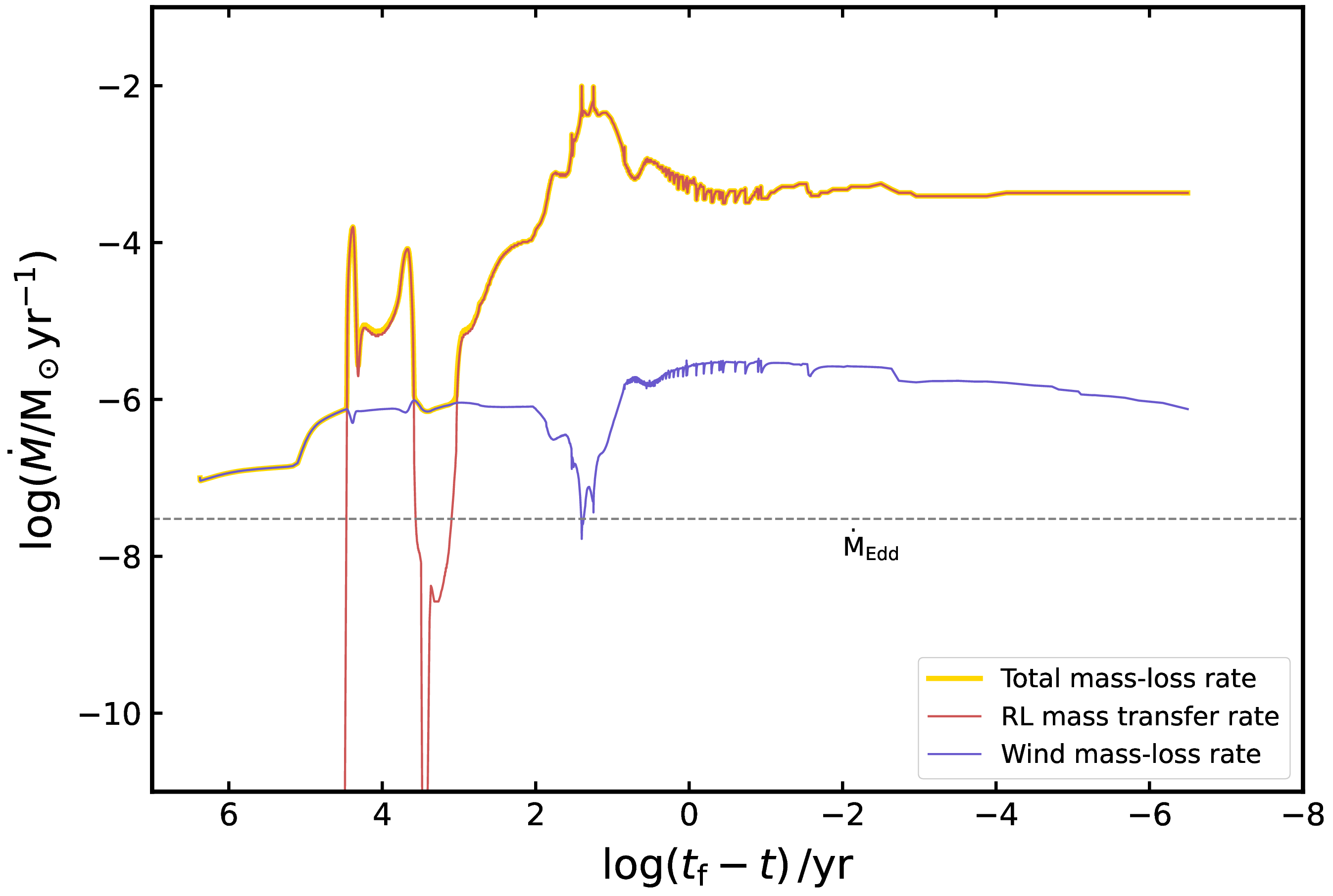}
	\caption{
		Mass-loss rate versus remaining time ($t_{\rm f}-t$) for the $2.67\,\rm M_\odot$ He star companion,
		in which $t$ and $t_{\rm f}$ are the stellar age and the total evolutionary time, respectively.
		The purple, red and yellow lines
		represent the stellar wind mass-loss rate, the mass-transfer rate and the total mass-loss rate, respectively.
		The dashed line indicates the Eddington accretion rate of the NS,
		i.e. $\dot M_{\rm Edd}=3 \times 10^{-8}\,\rm M_\odot \rm yr^{-1}$ for accretion of helium.}
\label{fig:mass-loss}
\end{figure}
\section{Results}

\begin{figure}
\centering\includegraphics[width=\columnwidth*3/3]{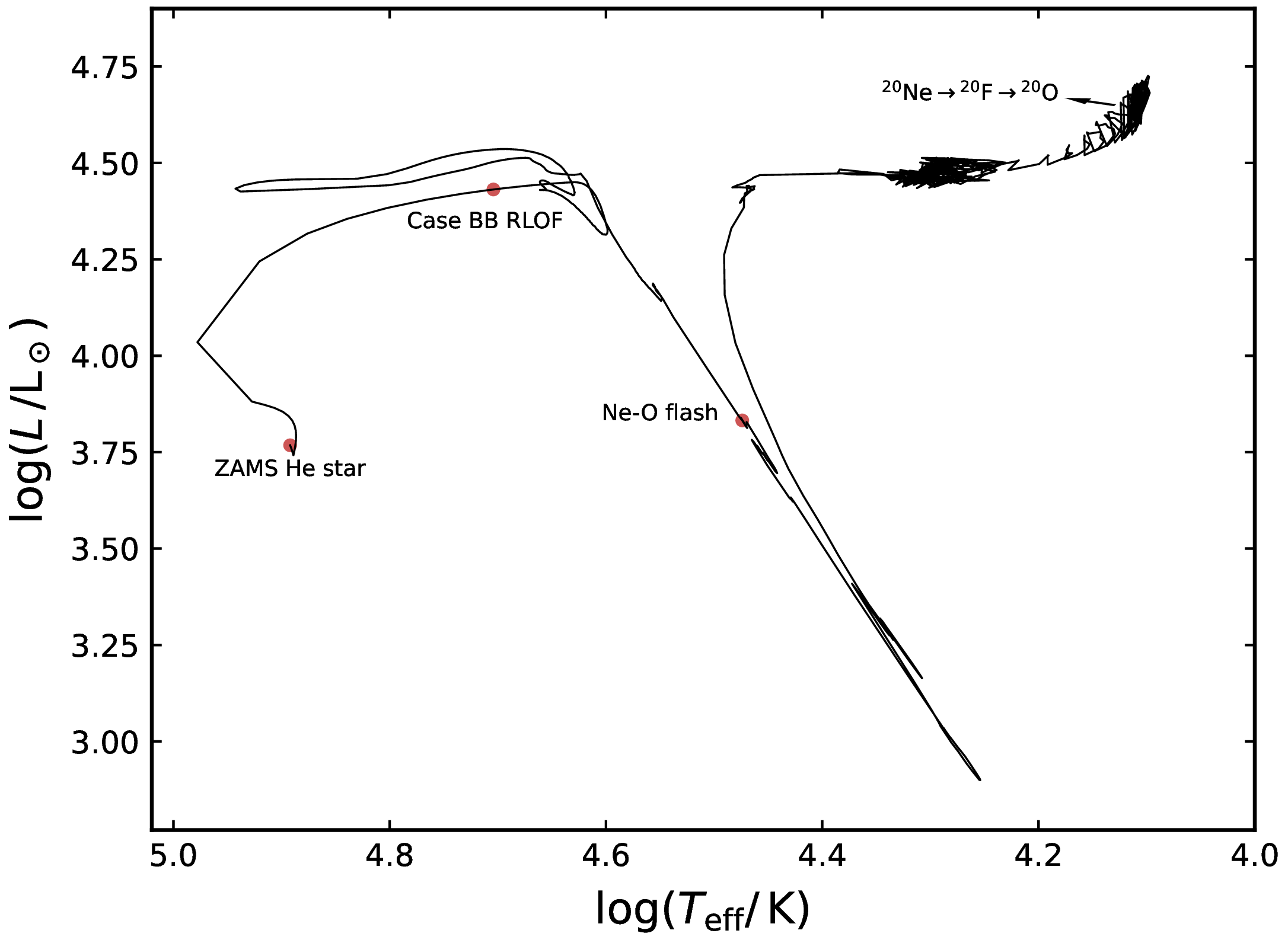}
\caption{H-R diagram of a $2.67\,\rm M_\odot$ He star companion from He-ZAMS to $e$-capture on $\rm^{20}Ne$,
	in which the brown dots denote the moment when the He star evolution, the Case BB RLOF,
	the Ne-O flashes and the $e$-capture on $\rm^{20}Ne$ begin.}
\label{fig:H-R}
\end{figure}

\subsection{Typical binary evolutionary example for EC-SNe}\label{3.1}
\begin{figure}
	\centering\includegraphics[width=\columnwidth*3/3]{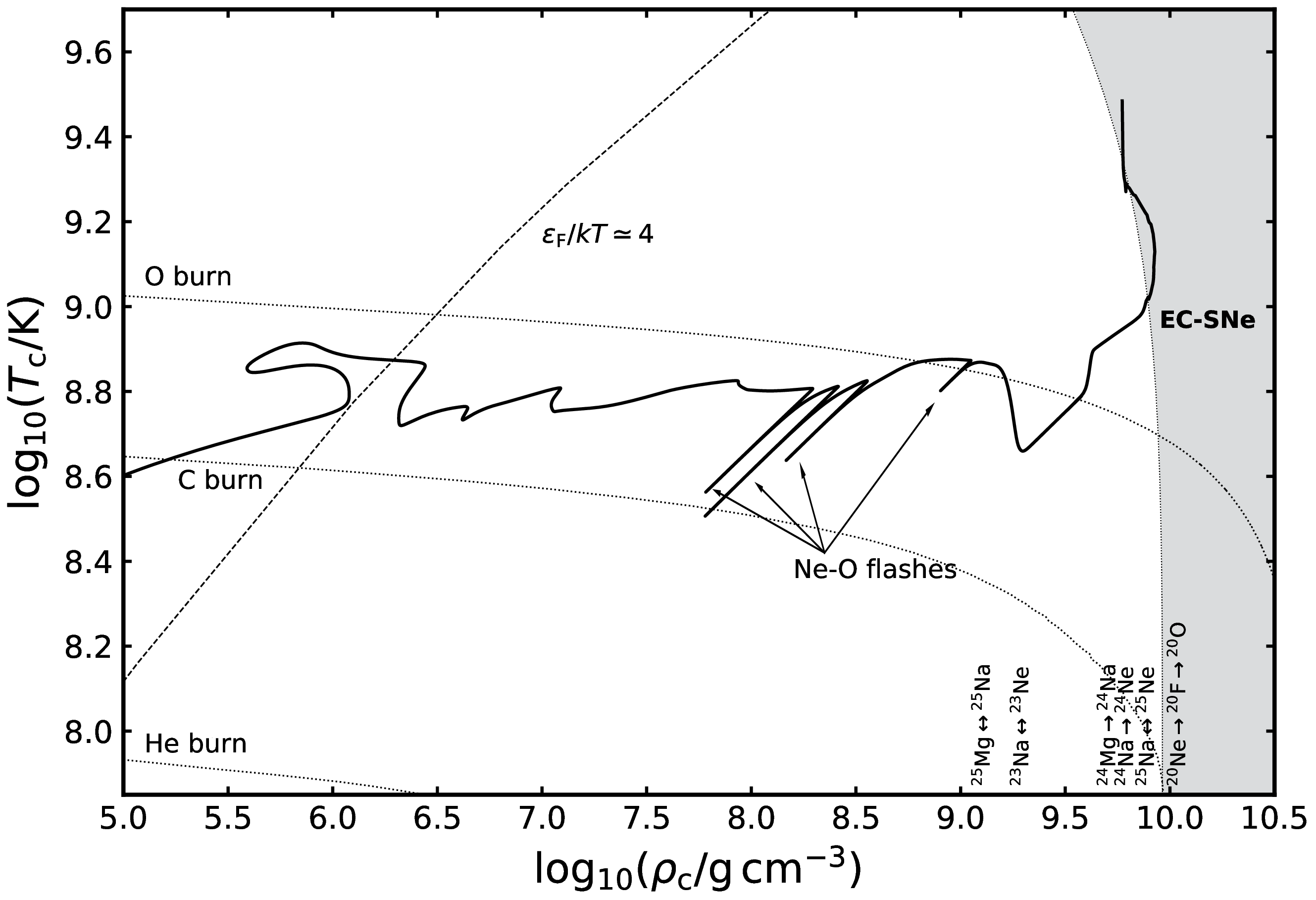}
	\caption{Central temperature versus central density for the $2.67\rm\,M_\odot$ He star companion.
		Grey shading denotes the region where the $e$-capture on $\rm^{20}Ne$ occurs.
		The dotted lines represent helium, carbon and oxygen burning ignition curves, respectively.
		The dashed line denotes the separation of degenerate and non-degenerate regions ($\epsilon_{\rm F}/k T \simeq 4$).}
	\label{fig:pc-tc1}
\end{figure}

\begin{figure}
	\centering\includegraphics[width=\columnwidth*3/3]{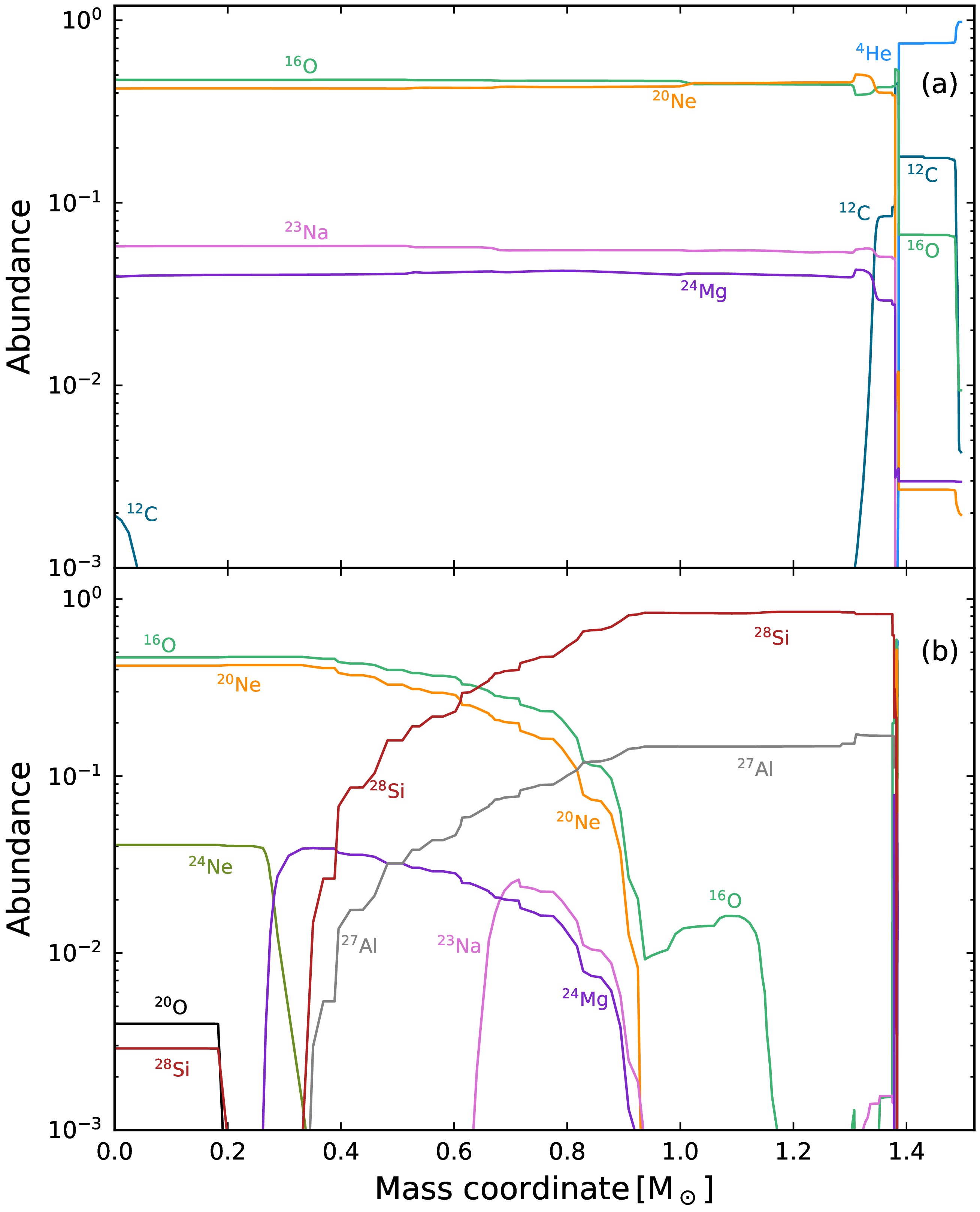}
	\caption{Panel (a): the chemical structure of the donor after carbon burning stage.
		At this moment, the donor mass is $\sim1.50\rm\,M_\odot$,
		and the donor consists a $\sim1.385\rm\,M_\odot$ ONe core
		and a $\sim0.115\rm\,M_\odot$ He envelope.
		Panel (b): the chemical structure of the donor
		when the $e$-capture reactions on $^{20}\rm Ne$ occur.
	At this moment, the donor consists of an ONe core and a thick Si shell.}
	\label{fig:ele-ob}
\end{figure}

In Figs\,\ref{fig:mass-loss}$-$\ref{fig:ele-ob},
we present an example of the evolution of a NS+He star binary
with $M_{\rm He}^{\rm i} = 2.67\rm\,M_\odot$ and $P_{\rm orb}^{\rm i}=1.0$\,d,
in which the final fate of the He star companion is an EC-SN.
Fig.\,\ref{fig:mass-loss} shows the evolution of mass-loss rate of He star companion as a function of remaining time
($t_{\rm f}-t$) until $e$-capture on $\rm^{20}Ne$ take place,
where $t$ and $t_{\rm f}$ are the stellar age and the total evolutionary time, respectively.
The total simulated stellar age form He-ZAMS to $e$-capture is $\sim 2.37$\,Myr.
After the central helium exhaustion,
the He star gradually expands,
thereby leading to the initiation of Roche-lobe overflow (so-called Case BB RLOF) at $t\sim2.35$\,Myr.
Fig.\,\ref{fig:H-R} shows the evolution of the companion in the H-R diagram.
The points marked by the brown dots along the evolutionary track correspond to He star ZAMS,
the onset of Case BB RLOF and other key processes.
Fig.\,\ref{fig:pc-tc1} illustrates the whole evolution of
central temperature ($T_{\rm c}$) and central density ($\rho_{\rm c}$) for the He star companion.

Following central helium and carbon burning stage,
the He star companion develops a
$\sim 1.385\,\rm M_\odot$ ONe core
which is surrounded by a $\sim 0.1\,\rm M_\odot$ He-layer.
Fig.\,\ref{fig:ele-ob}a shows the chemical structure of the mass donor
after the carbon burning stage.
At this moment,
the ONe core is mainly composed of oxygen, neon, sodium and magnesium.
Subsequently, neon is ignited off-center at mass coordinates ($M_{\rm c}$) of $\sim 0.8\,\rm M_\odot$
due to the temperature inversion caused by the neutrino emission in the center of core (i.e. Ne-shell flash).
From Fig.\,\ref{fig:pc-tc1},
it can be seen that the center reaches higher density and temperature
after each Ne-shell flash quenches
\citep[][]{2013ApJ...772..150J}.
Following a few Ne-shell flashes,
the central density increases to $\rm log_{10}(\rho_c/g\,cm^{-3})\approx9.1$,
resulting in the initiation of Urca processes.
The central temperature decreases because of the Urca reactions.
Meanwhile,
the Urca reactions accelerate the contraction of the metal core,
thereby leading to higher central density.

As the central density gradually increases,
$e$-capture on $^{24}\rm Mg$ and $^{20}\rm Ne$ take place at
$\rm log_{10}(\rho_c/g\,cm^{-3})\approx9.6$ and
$\rm log_{10}(\rho_c/g\,cm^{-3})\approx9.9$, respectively (see Fig.\,\ref{fig:pc-tc1}).
Fig.\,\ref{fig:ele-ob}b shows the chemical structure of the mass donor
when the $e$-capture on $\rm^{20}Ne$ occurs.
We can see that the isotope $^{20}\rm O$ is produced in the center owing to
the reactions of $^{20}\rm Ne$$(e^-,\nu_e)^{20}\rm F$$(e^-,\nu_e)^{20}\rm O$.
Meanwhile,
a thick Si-rich mantle is formed because of the Ne-shell flashes.
At this moment,
the final mass of the He envelope is $\sim8.2\times10^{-4}\rm\,M_\odot$,
thus the He lines are not visible in the spectra
after the core-collapse event
\citep[][]{2012MNRAS.422...70H}.

\subsection{Parameter space for EC-SNe}
By calculating a large number of NS+He star systems with different $M_{\rm He}^{\rm i}$ and $P_{\rm orb}^{\rm i}$,
we obtained the initial contours for producing EC-SNe.
Table\,\ref{table:1} lists the information about the evolution of NS+He star binaries
($Z=0.02$) at the boundaries of the parameter space for EC-SNe,
in which the selected models correspond to the lowest simulated masses
	(for the given initial orbital period) to produce either Fe CC-SNe or EC-SNe.
The metal core mass for EC-SNe ranges from $\sim 1.385\rm\,M_\odot$ to $1.43\rm\,M_\odot$.
If the final core mass is lower than $\sim 1.385\rm\,M_\odot$,
then the explosive oxygen burning can be triggered
owing to the convective Urca process,
resulting in that the He star companions explode as SNe Ia eventually
\citepalias[see][]{2023MNRAS.526..932G}.
If the final core mass is larger than $\sim 1.43\rm\,M_\odot$,
then the final fates of He star companions are Fe CC-SNe.

\begin{figure}
	\centering\includegraphics[width=\columnwidth*3/3]{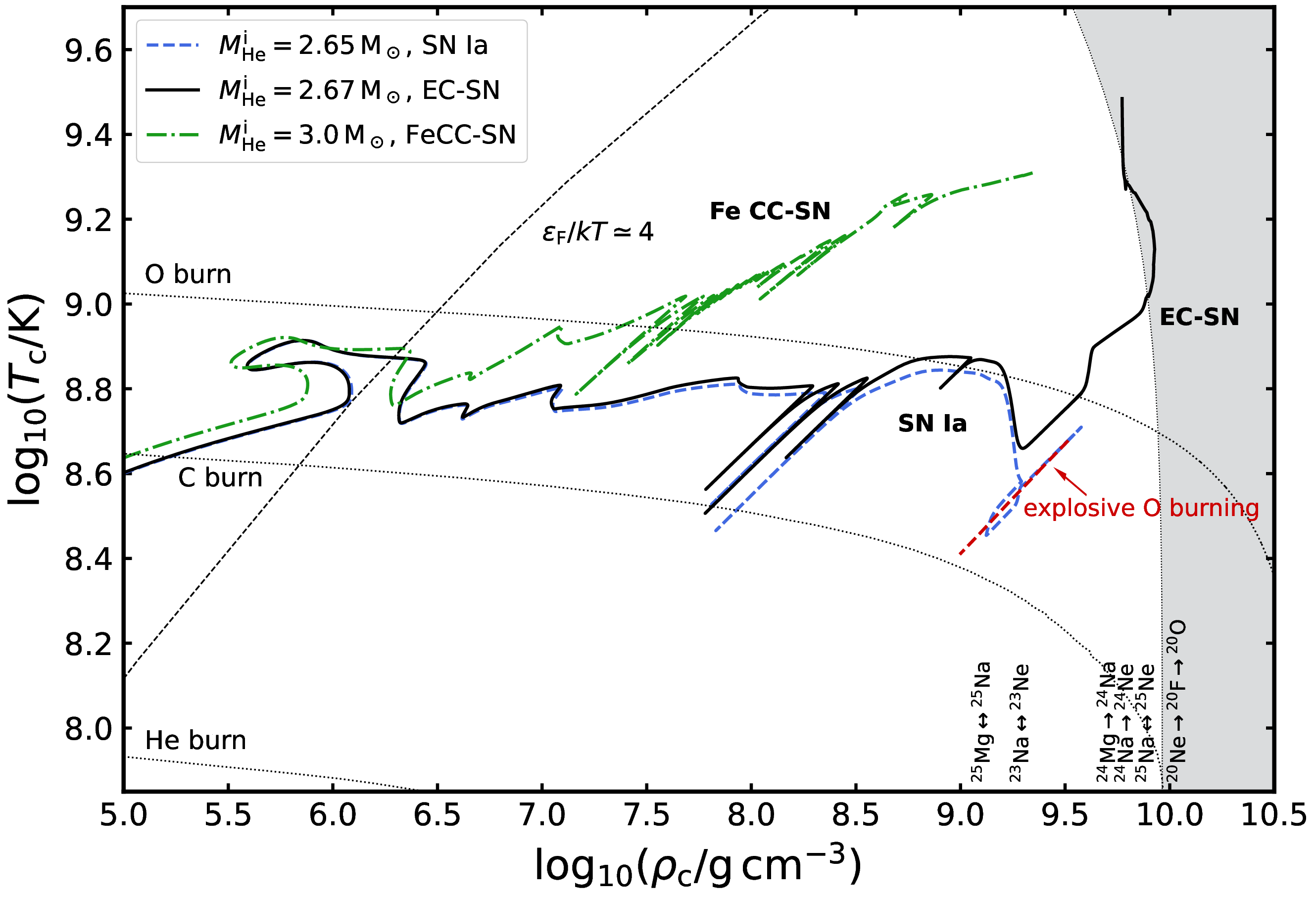}
	\caption{Evolutionary track of the central temperature and the central density
		for NS+naked He star binaries with different initial He star masses
		(i.e. $M_{\rm He}^{\rm i}=2.65, 2.67$ and $3.0\rm\,M_\odot$),
		in which the initial orbital period of all models is $1.0$\,d.}
	\label{fig:pc-tc}
\end{figure}

\begin{figure}
	\centering\includegraphics[width=\columnwidth*3/3]{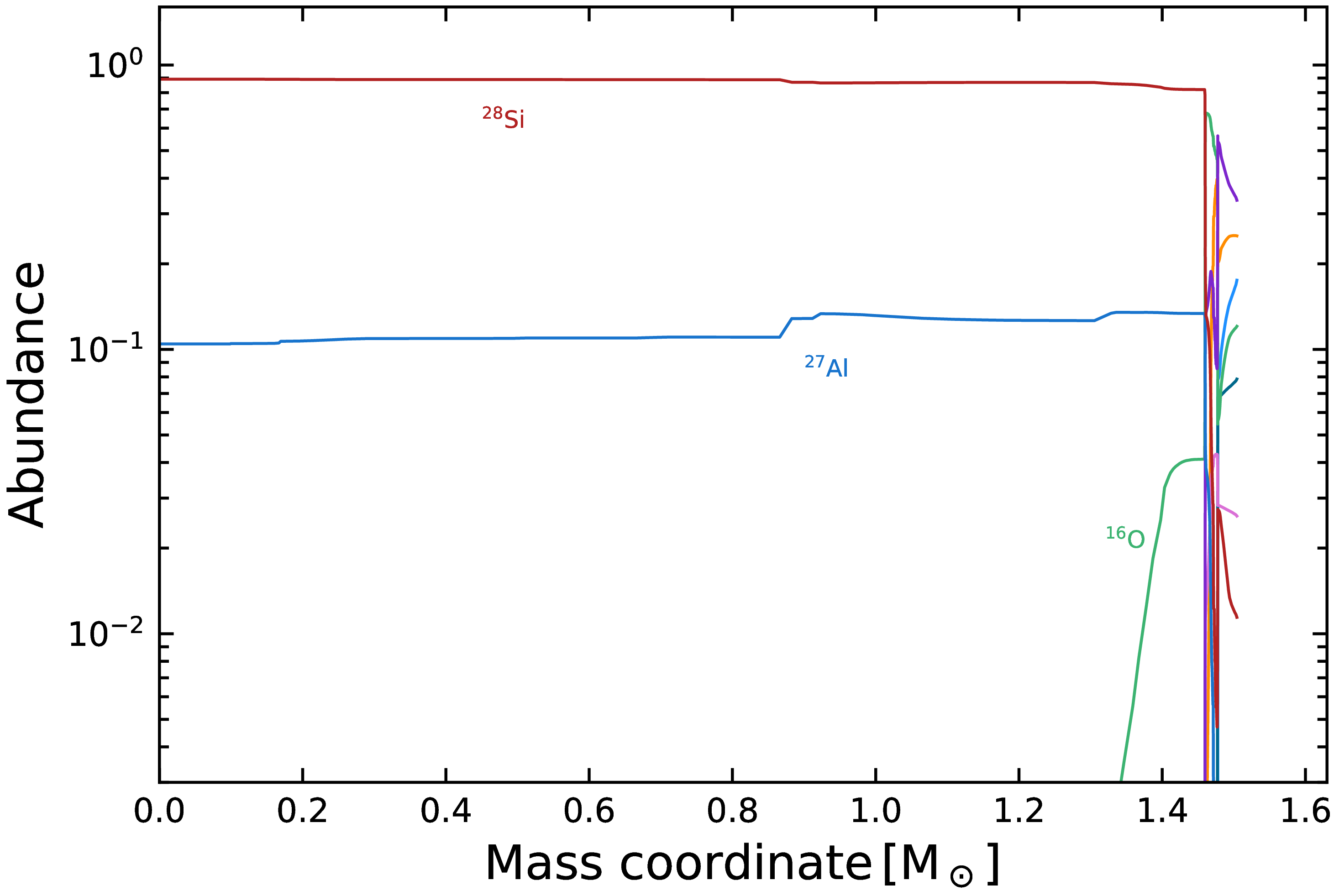}
	\caption{Final chemical structure for the He star companion with initial mass of $3.0\rm\,M_\odot$,
		in which the final core mass is $\sim1.55\rm\,M_\odot$ and its final fate is an Fe CC-SN.}
\label{fig:ele-fe}
\end{figure}

Fig.\,\ref{fig:pc-tc} shows $T_{\rm c}$ versus $\rho_{\rm c}$ for three representative NS+He star binaries
which terminate their evolution as an SN Ia, an EC-SN and an Fe CC-SN, respectively.
For the model with $M^{\rm i}_{\rm He}=2.65\,\rm M_\odot$,
the He star companion develops a $\sim 1.378\,\rm M_\odot$ ONe core,
and the explosive oxygen burning takes place
owing to the convective Urca process,
resulting in the formation of an SN Ia
\citepalias[see][]{2023MNRAS.526..932G}.
For the model with $M^{\rm i}_{\rm He}=2.67\,\rm M_\odot$,
the He star companion develops a $\sim 1.385\,\rm M_\odot$ ONe core,
and the He star will undergo EC-SN after $e$-capture on $\rm^{20}Ne$ occurs.
For the model with $M^{\rm i}_{\rm He}=3.0\,\rm M_\odot$,
the He star companion develops a $\sim 1.55\,\rm M_\odot$ ONe core,
and the Ne-shell flash takes place at $M_{\rm c}\approx0.3\,\rm M_\odot$,
which is closer to the center than the model with a $2.67\,\rm M_\odot$ He star companion.
Fig.\,\ref{fig:ele-fe} shows the final chemical structure of the He star companion.
The Ne-O burning flame can propagate to the center,
leading to the formation of a silicon core, and thereby an iron core-collapse.

\begin{figure}
	\centering\includegraphics[width=\columnwidth*3/3]{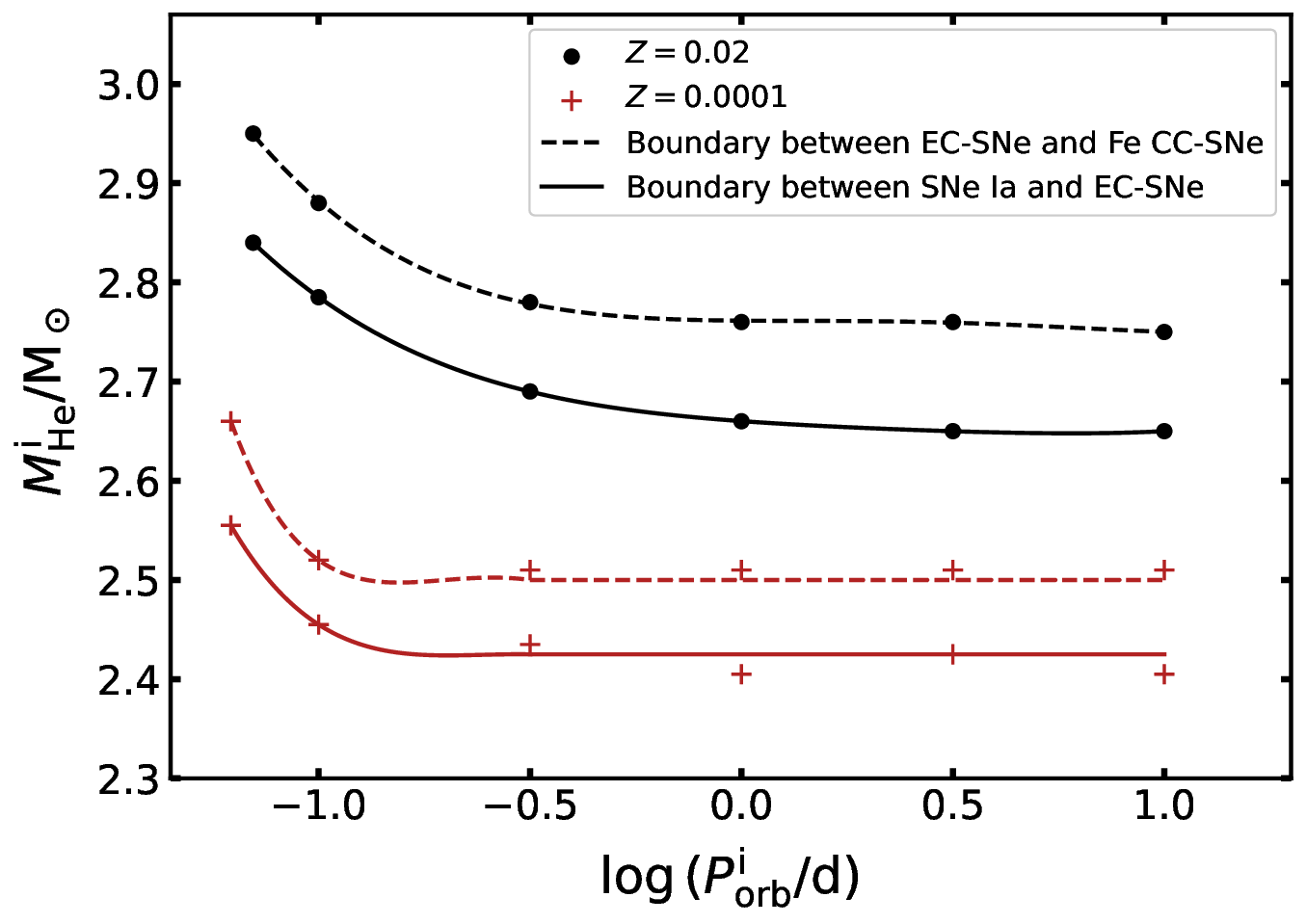}
	\caption{Initial parameter spaces in the
		$\rm log\,$$P^{\rm i}_{\rm orb}-M^{\rm i}_{\rm He}$ plane for NS+naked He star binaries
		that can produce EC-SNe.
		The dashed lines represent the boundaries between EC-SNe and Fe CC-SNe,
		and the solid lines represent the boundaries between SNe Ia and EC-SNe.
	}
	\label{fig:ECSN}
\end{figure}

Fig.\,\ref{fig:ECSN} represents the initial parameter spaces
for EC-SNe in the log$P_{\rm orb}^{\rm i}-M_{\rm He}^{\rm i}$ diagram with two metallicities (i.e. $Z=0.0001, 0.02$),
from which the metallicity has a strong influence on the parameter spaces.
The left boundaries of the parameter spaces are set by the condition
that RLOF takes place when the He star companions are on He-ZAMS stage.
If the initial orbital period is $\gtrsim10$\,d,
then the He star companions have a similar final fate.
The dashed lines represent the boundaries between EC-SNe and Fe CC-SNe,
and the solid lines represent the boundaries between SNe Ia and EC-SNe.
We found that for a lower metallicity,
the parameter spaces shifted to low-mass He star companion.
This is because lower metallicities result in weaker stellar wind mass-loss
and larger metal core for a given initial He star mass
\citep[e.g.][]{2022A&A...668A.106C,2023A&A...671A.134A}.
Meanwhile, lower metallicities lead to smaller radii of He ZAMS stars,
thus the left boundaries of the parameter space shifts slightly to shorter initial period
\citep[see also][]{2010A&A...515A..88W}. 

\begin{table*}
	\centering
	
	\caption{
		Evolutionary properties of the NS+He star binaries ($Z=0.02$) at boundaries of the parameter space for EC-SNe,
		where the selected models correspond to the lowest simulated masses
		(for the given initial orbital period) to produce either Fe CC-SNe or EC-SNe.
		$M_{\rm He}^{\rm i}$ and log$P_{\rm orb}^{\rm i}$ are the initial mass of the He star companion
		and the initial orbital period;
		$M_{\rm core, f}$ and $M_{\rm env, f}$ are the final metal core mass and the remaining helium envelope mass of the donor prior to SN;
		$\Delta M_{\rm NS}$ is the accreted mass of NS;
		$P_{\rm spin}^{\rm min}$ is the minimum spin period of NS;
		$P_{\rm orb}^{\rm pre}$ is the final orbital period prior to SN;
		$t_{\rm f}$ is the total evolution time of He star companion;
		the last column is the evolutionary outcome of the He star companion.
	}
	\label{table:1}
	\begin{tabular}{ c  c c c ccc c c c c c }
		\toprule
		\hline 
		Set		&&$M_{\rm He}^{\rm i}$  &log$P_{\rm orb}^{\rm i}$  &$M_{\rm core, f}$	&$M_{\rm env, f}$ & $\Delta M_{\rm NS}$  &$P_{\rm spin}^{\rm min}$ &$P_{\rm orb}^{\rm pre}$  &$t_{\rm f}$	&Final fate\\
		&&($\rm M_\odot$)&   (d)& ($\rm M_\odot$)& ($\rm M_\odot$) & $(\rm M_\odot)$   & (ms) &(d)	&(Myr) &\\
		\hline 
		1	&&$2.95$		&$-1.15$	&$1.431$	&4.5e-3		&1.8e-3			&$38.7$	&$0.069$	&$2.02$	&Fe CC-SN\\
		2	&&$2.84$		&$-1.15$	&$1.381$	&3.0e-4		&2.0e-3		&$36.0$	&$0.075$ 	&$2.14$	&EC-SN\\
		\hline 
		3	&&		$2.88$		&$-1.00$	&$1.431$	&3.0e-3		&1.6e-3	 &$43.3$	&$0.107$	&$2.10$	&Fe CC-SN\\
		4	&&		$2.79$ 		&$-1.00$	&$1.389$	&5.0e-4		&1.7e-3  &$40.6$	&$0.113$	&$2.20$	&EC-SN\\
		\hline 
		7	&&		$2.78$		&$-0.50$	&$1.430$	&4.7e-3		&9.0e-4	 &$65.7$	&$0.361$	&$2.22$ &Fe CC-SN\\
		6	&&		$2.70$		&$-0.50$	&$1.396$	&7.3e-4		&8.7e-4	 &$67.1$	&$0.378$	&$2.32$	&EC-SN\\
		\hline
		7	&&		$2.77$		&$0$		&$1.436$	&3.0e-3		&7.1e-4		 &$78.2$	&$1.153$	&$2.23$	&Fe CC-SN\\
		8	&&		$2.67$		&$0$		&$1.385$	&6.3e-4		&8.2e-4  &$70.2$	&$1.226$	&$2.37$	&EC-SN\\
		\hline
		9	&&		$2.76$		&$0.5$		&$1.435$	&5.4e-3		&5.7e-4	 &$92.2$	&$3.676$	&$2.25$	&Fe CC-SN\\
		10	&&		$2.65$		&$0.5$		&$1.383$	&4.5e-4		&6.7e-4	 &$81.6$	&$3.915$	&$2.40$	&EC-SN\\	
		\hline
		11	&&		$2.76$		&$1$		&$1.433$	&6.4e-3		&3.3e-4	 &$138.3$	&$11.65$	&$2.25$	&Fe CC-SN\\
		12	&&		$2.66$		&$1$		&$1.384$	&2.5e-4		&4.8e-4	 &$104.8$	&$12.39$	&$2.38$	&EC-SN\\	
		\hline
		13$^{\rm a}$	&&		$2.73$		&$-1.15$	&$1.433$	&1.6e-3		&2.0e-3		 &$36.8$	&$0.057$	&$2.23$	&Fe CC-SN\\
		14$^{\rm a}$	&&		$2.64$		&$-1.15$	&$1.385$	&4.2e-4		&2.1e-3	 &$34.2$	&$0.062$	&$2.35$	&EC-SN\\
		\hline
		15$^{\rm a}$	&&		$2.54$		&$1$		&$1.443$	&1.7e-2		&2.0e-4		 &$208.5$	&$9.502$	&$2.56$	&Fe CC-SN\\
		16$^{\rm a}$	&&		$2.49$		&$1$		&$1.393$	&7.0e-4		&3.0e-4		 &$157.1$	&$10.10$	&$2.66$	&EC-SN\\
		\hline
	\end{tabular}
	\begin{tablenotes}
		\item[]$^{\rm a}$ Models with a stellar wind efficiency of $0.1$.
	\end{tablenotes}
\end{table*}
\section{Double neutron star systems}
The simulations show that
the He star companions in NS binaries may explode as SNe Ia
if the degenerate ONe core masses
range from $\sim 1.335-1.385\,\rm M_\odot$
\citepalias[see][]{2023MNRAS.526..932G}.
Meanwhile, the NSs spin up during the mass-accretion phase,
resulting in the formation of isolated mildly recycled pulsars
after the He star companions experience SN Ia explosions.
If the ONe core masses are larger than $\sim 1.385\,\rm M_\odot$,
then the He star companions collapse into NSs through EC-SNe or Fe CC-SNe,
which may result in the formation of DNS systems
\citep[e.g.][]{2017ApJ...846..170T,2018MNRAS.481.1908K}.
Accordingly,
we will study the DNS systems originating from the EC-SN channel,
and compare our calculation results with observations.

\subsection{The effects of SN explosions}
It has been suggested that the eccentricity ($e$) and the orbital period of NS binaries 
will be changed after the companion star undergoes a SN explosion,
due to the sudden mass loss and the kick velocity ($V_{\rm k}$)
added to the newly-born NS.
The orientation of the NS kick velocity is controlled by two angles,
i.e. the angle between $V_{\rm k}$ and the pre-SN orbital velocity ($0\leq\theta\leq180^{\circ}$),
and the positional angle of $V_{\rm k}$ out of the orbital plane ($-180^{\circ}<\phi\leq180^{\circ}$).
Thus, the ratio of pre-SN orbital separation ($a_0$)
to the post-SN orbital separation ($a$) can be given by
\citep[e.g.][]{1983ApJ...267..322H, 2003MNRAS.344..629D, 2016ApJ...816...45S}
\begin{equation}
	\frac{a_0}{a} = 2 - \frac {M_0}{M_0-\Delta M}(1+\nu^2+2\nu\,\rm cos\,\theta),
\end{equation}
where $M_0$ and $\Delta M$ are the pre-SN total mass,
and the ejected mass from the exploding star, respectively.
In addition,
we defined $\nu = V_{\rm k}/V_0$,
in which $V_0$ is relative velocity between the two stars.
The eccentricity after SN explosions can be written as
\begin{equation}
	1-e^2 = \frac{a_0}{a}\frac {M_0}{M_0-\Delta M}[1+2\nu\,\rm cos\,\theta
	+\nu^2(\rm cos^2\theta+sin^2\theta\,sin^2\phi)].
\end{equation}
\begin{figure}
	\centering\includegraphics[width=\columnwidth*3/3]{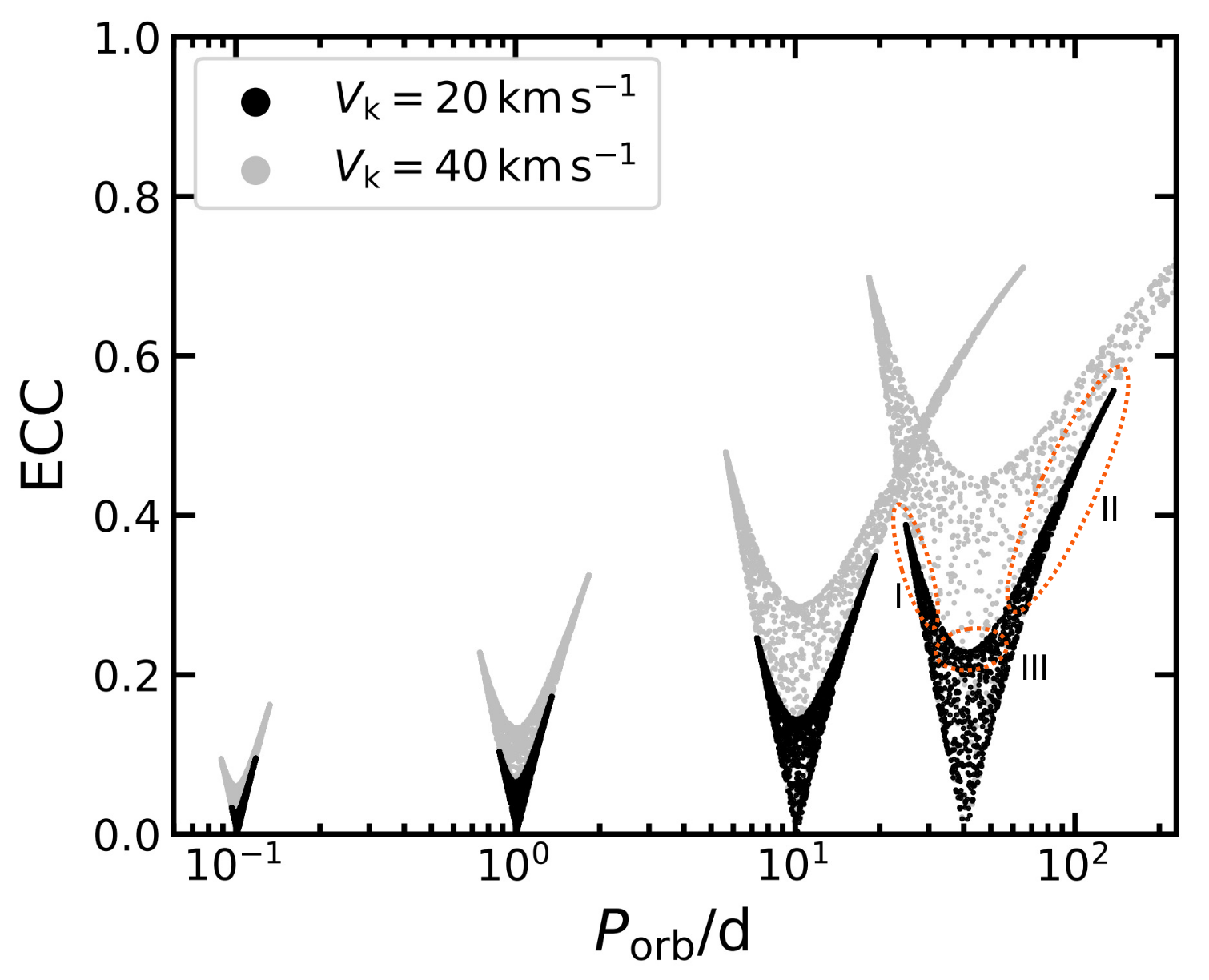}
	\caption{Distribution of the simulated post-SN systems in the $P_{\rm orb}-e$ diagram,
		in which we performed $2000$ cycles with randomly oriented NS kicks for each pre-SN system.
		We adopted four pre-SN orbital periods (i.e. $0.1$, $1$, $10$ and $40$\,d)
		and two NS kicks (i.e. $V_{\rm k}=20$ and $40\,\rm km\,s^{-1}$).
		Regions I, II and III enclosed by red dotted ovals
			are the parameter spaces that the post-SN systems have relatively high probabilities.}
	\label{fig:kick-pe}
\end{figure}

\begin{figure*}
	\centering\includegraphics[width=\columnwidth*9/5]{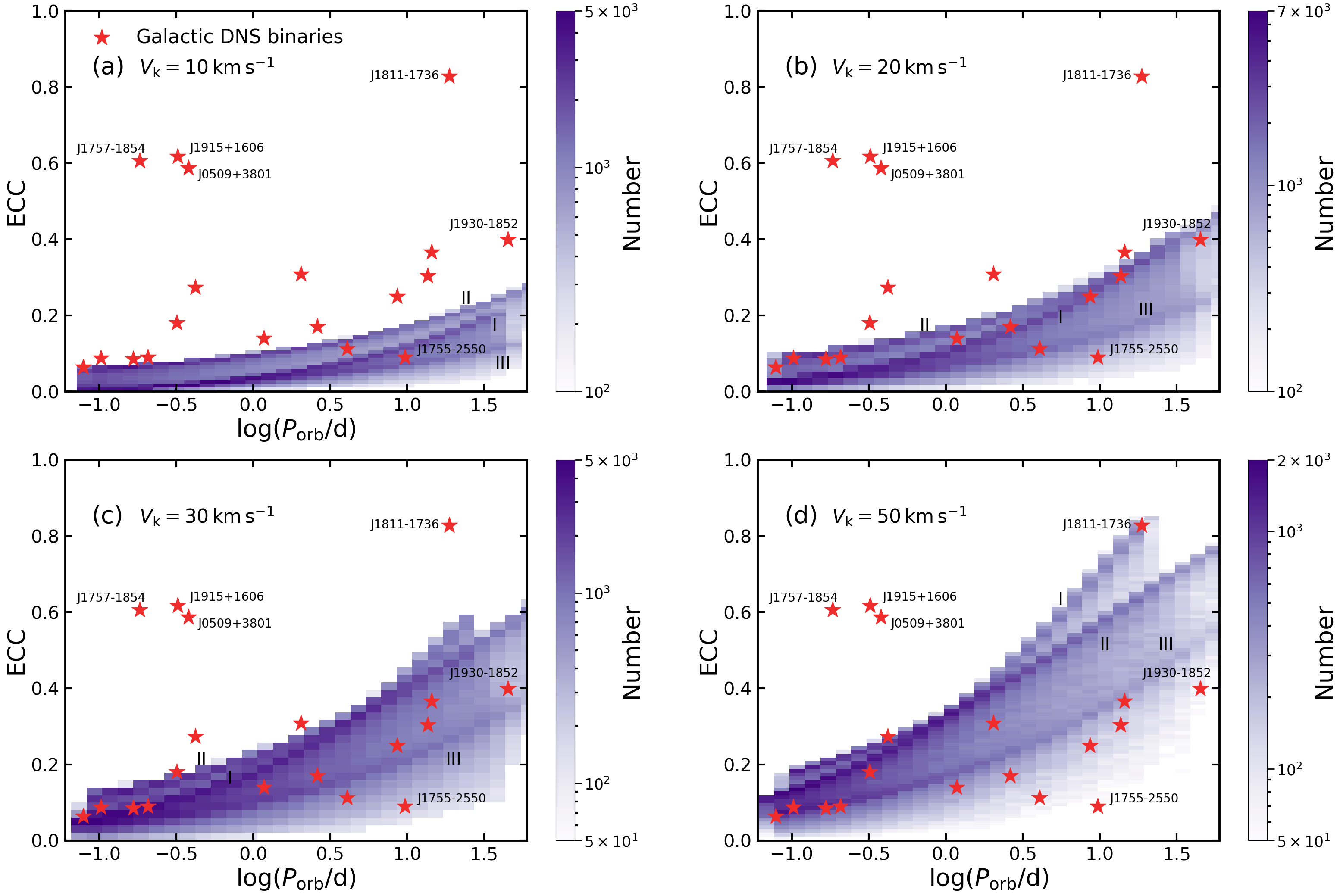}
	\caption{Distribution of the simulated post-SN systems in the $P_{\rm orb}-e$ diagram (the purple regions),
		in which we set $500$ pre-SN $P_{\rm orb}$ that vary logarithmically from $0.07$\,d to $50$\,d and the NS kicks
		are in the range of $10-50\,\rm km\,s^{-1}$.
We calculated $1000$ cycles with randomly oriented NS kicks for each pre-SN system.
Lines I, II and III are the regions where post-SN systems are located
with relatively high probability, corresponding to
$\theta\sim180^{\circ}$, $\theta\sim0$ and $\phi\sim0, \pm180^{\circ}$, respectively.
		The red stars are the Galactic DNS systems in observations.
		The observed data of DNS binaries are taken from the ATNF Pulsar Catalogue, \href{http://www.atnf.csiro.au/research/pulsar/psrcat}{http://www.atnf.csiro.au/research/pulsar/psrcat}
		\citep[version 1.71, November 2023;][]{2005AJ....129.1993M}.}
	\label{fig:pe}
\end{figure*}

We assume that the NS mass is $1.35\,\rm M_\odot$.
In our simulations,
the final companion star mass prior to SN ranges from $\sim1.385\,\rm M_\odot$ to $1.43\,\rm M_\odot$,
thus we adopt an average value of $1.4\,\rm M_\odot$. 
Accordingly, the pre-SN total mass is $2.75\,\rm M_\odot$.
The ejected mass for EC-SNe is still highly uncertain.
Previous studies generally suggested that the collapse of these objects can produce
small ejected mass from $0.01-0.2\,\rm M_\odot$
\citep[e.g.][]{1996ApJ...457..834T, 2006A&A...450..345K, 2012ApJ...749...91F}.
In this work,
the value of $\Delta M$ is set to be $0.08\,\rm M_\odot$
\citep[][]{2018ApJ...867..124S}.
The small value of $\Delta M$ means that the properties of post-SN binaries
are dominated by the natal NS kicks.
In addition,
it has been suggested that EC-SNe may produce low NS kicks
owing to the fast explosions and relatively low explosion energies,
while the value of NS kicks is still under debate
\citep[e.g.][]{2004ApJ...612.1044P, 2006A&A...450..345K, 2017A&A...608A..57V, 2017ApJ...837...84J, 2018ApJ...865...61G, 2020MNRAS.496.2039S}.
\citet{2015MNRAS.451.2123T} suggested that
the amount of ejecta is extremely small for EC-SNe and low-mass Fe CC-SNe,
resulting in small NS kicks less than $\sim10\,\rm km\,s^{-1}$.
In addition,
simulations of EC-SNe show that the explosion energy is about $10^{50}$\,erg,
which leads to the kick velocities lower than $\sim50\,\rm km\,s^{-1}$
\citep[e.g.][]{2006A&A...450..345K, 2006ApJ...644.1063D, 2016MNRAS.456.4089B, 2017ApJ...837...84J}.
Accordingly,
we assume that $V_{\rm k}\leq50\,\rm km\,s^{-1}$ for EC-SNe.
%
\subsection{DNS systems in the $P_{\rm orb}-e$ diagram}

Fig.\,\ref{fig:kick-pe} shows the effect of NS kicks
and pre-SN orbital period on the distribution of post-SN systems in the $P_{\rm orb}-e$ diagram,
where we performed $2000$ cycles with randomly oriented NS kicks for each pre-SN system.
Obviously, larger $V_{\rm k}$ and longer pre-SN orbital period
easily lead to higher post-SN eccentricity and longer post-SN $P_{\rm orb}$.
We note that if the systems have short pre-SN $P_{\rm orb}$ ($\lesssim1$\,d)
and relatively low $V_{\rm k}$,
then the post-SN $P_{\rm orb}$ does not change significantly.
Additionally,
the binaries may be disrupted if they have wide-orbits.
For example, if a pre-SN binary has parameters with
($V_{\rm k}, P_{\rm orb})=(40\,\rm km\,s^{-1}, 40\,d)$,
then the disruption probability for this system is $32\%$.
Moreover,
we found that the probability of the post-SN system
located in the minimum and maximum period is relatively high,
corresponding to $\theta$ around $180^{\circ}$ and $0$, respectively.

\begin{table*}
	\centering
	\caption{Properties of Galactic DNS systems.
		$M_{\rm total}$, $M_{\rm psr}$, $M_{\rm comp}$ are the binary total mass, the mass of radio pulsar and the mass of companion;
		$P_{\rm orb}$, $e$, $P_{\rm spin}$ and $\dot P_{\rm spin}$ are the binary orbital period, the eccentricity of DNS system, the spin period of pulsar and its derivative;
		$V_{\rm k}$ and $P_{\rm orb}^{\rm pre}$ are the NS kick velocity and pre-SN orbital period estimated in this work.}
	\label{table:2}
	\begin{tabular}{ c  c c c ccc c c c c l }
		\toprule
		\hline
		&&	$M_{\rm total}$	&$M_{\rm psr}$	&$M_{\rm comp}$	&$P_{\rm orb}$	&$e$	&$P_{\rm spin}$	&$\dot P_{\rm spin}$
		&$V_{\rm k}$	&$P_{\rm orb}^{\rm pre}$\\
		\hline 
		Pulsar name	&&	($\rm M_\odot$)	&($\rm M_\odot$)	&($\rm M_\odot$)	&(d)	& &	(ms)	&($10^{-18}$)&($\rm km\,s^{-1}$)	&(d)\\
		\hline
		$\rm J0453+1559^1$&&	$2.734$	&$1.559$	&$1.174$	&$4.072$	&$0.113$	&$45.8$	&$0.186$	&$\lesssim 80$	&$\sim3-5$\\
		$\rm J0509+3801^{2}$&&	$2.805$	&$1.348$	&$1.468$	&$0.380$	&$0.586$	&$76.5$	&$80.54$	&$\gtrsim 150$	&$\lesssim0.9$\\
		$\rm J0737-3039^3$&&	$2.587$	&$1.338$	&$1.249$	&$0.102$	&$0.088$	&$22.7$	&$1.76$		&$\sim 20-200$	&$\sim0.08-0.1$\\
		$\rm J1411+2551^4$&&	$2.538$	&$\textless 1.64$ &$\textgreater 0.92$	&$2.616$ &$0.170$ &$62.5$ &$0.096$ &$\lesssim 100$	&$\sim2-4$\\
		$\rm J1518+4904^5$&&	$2.718$	&$1.41$		&$1.31$		&$8.634$ 	&$0.249$	&$40.9$	&$0.027$	&$\sim 20-80$	&$\sim4-15$\\
		$\rm J1537+1155^{6}$&&	$2.678$	&$1.333$	&$1.346$	&$0.421$	&$0.274$	&$37.9$	&$2.42$		&$\sim30-200$	&$\sim0.2-0.7$\\
		$\rm J1753-2240^7$&&	...		&...		&...		&$13.64$	&$0.304$	&$95.1$	&$0.970$	&$\sim 20-80$	&$\sim7-20$\\
		$\rm J1755-2550^8$&&	...		&...		&...		&$9.696$	&$0.089$	&$315$	&$2434$		&$\lesssim 30$	&$\sim8-12$\\
		$\rm J1756-2251^9$&&	$2.569$ &$1.341$	&$1.230$	&$0.320$	&$0.181$	&$28.5$	&$1.018$	&$\sim 30-150$	&$\sim0.2-0.5$\\
		$\rm J1757-1854^{10}$&&	$2.733$	&$1.338$	&$1.395$	&$0.184$	&$0.606$	&$21.5$	&$2.630$	&$\gtrsim 150$	&$\lesssim0.3$\\
		$\rm J1759+5036^{11}$&&	$2.62$	&$\textless 1.92$ &$\textgreater 0.70$ &$2.043$	&$0.308$  &$176$  &$0.243$ &$\sim 30-150$	&$\sim1-4$\\
		$\rm J1811-1736^{12}$&&	$2.57$	&$\textless 1.75$ &$\textgreater 0.91$  &$18.78$ &$0.828$ &$104$ &$0.901$  &$\sim 80-300$	&$\sim1-50$\\
		$\rm J1829+2456^{13}$&&	$2.59$	&$\textless 1.38$ &$\textgreater 1.22$	&$1.176$ &$0.139$ &$41.0$&$0.053$  &$\sim 20-100$	&$\sim0.8-1.5$\\
		$\rm J1901+0658^{14}$&& $2.79$ &$\textless 1.68$ &$\textgreater 1.11$  &$14.45$ &$0.366$ &$75.7$&$0.217$&$$&\\
		$\rm J1906+0746^{15}$&&	$2.613$	&$1.291$	&$1.322$	&$0.166$	&$0.085$	 &$144.1$&$20300$	&$\sim 20-150$	&$\sim0.13-0.2$\\
		$\rm J1913+1102^{16}$&&	$2.89$	&$1.65$ &$1.24$	&$0.206$ &$0.090$&$27.3$ &$0.161$  &$\sim 20-150$	&$\sim0.15-0.3$\\
		$\rm J1915+1606^{17}$&&	...		&...		&...		&$0.323$	&$0.617$	&$59.03$ &$8.618$	&$\gtrsim 150$	&$\lesssim1.0$\\
		$\rm J1930-1852^{18}$&& ...		&...		&...		&$45.06$	&$0.400$	&$185.5$ &$19.00$	&$\sim 20-80$	&$\sim 15-100$\\
		$\rm J1946+2052^{19}$&& $2.50$	&$\textless 1.35$ &$\textgreater 1.17$	&$0.078$ &$0.064$ &$17.0$&$0.92$   &$\lesssim 150$	&$\sim 0.07-0.09$\\
		\hline
	\end{tabular}
	\begin{tablenotes}
		\item[]References. 1: \citet{2015ApJ...812..143M}; 2: \citet{2018ApJ...859...93L}; 3: \citet{2006Sci...314...97K}; 4: \citet{2017ApJ...851L..29M};
		5: \citet{2008A&A...490..753J}; 6: \citet{2014ApJ...787...82F}; 7: \citet{2009MNRAS.393..623K}; 8: \citet{2015MNRAS.450.2922N};
		9: \citet{2014MNRAS.443.2183F}; 10: \citet{2018MNRAS.475L..57C}; 11: \citet{2021ApJ...922...35A}; 12: \citet{2007A&A...462..703C};
		13: \citet{2005MNRAS.363..929C}; 14: \citet{2024MNRAS.tmp..903S};15: \citet{2015ApJ...798..118V}; 16: \citet{2018IAUS..337..146F};
		17: \citet{2010ApJ...722.1030W}; 18: \citet{2015ApJ...805..156S}; 19: \citet{2018ApJ...854L..22S}.
	\end{tablenotes}
\end{table*}

\begin{figure*}
	\centering\includegraphics[width=\columnwidth*9/5]{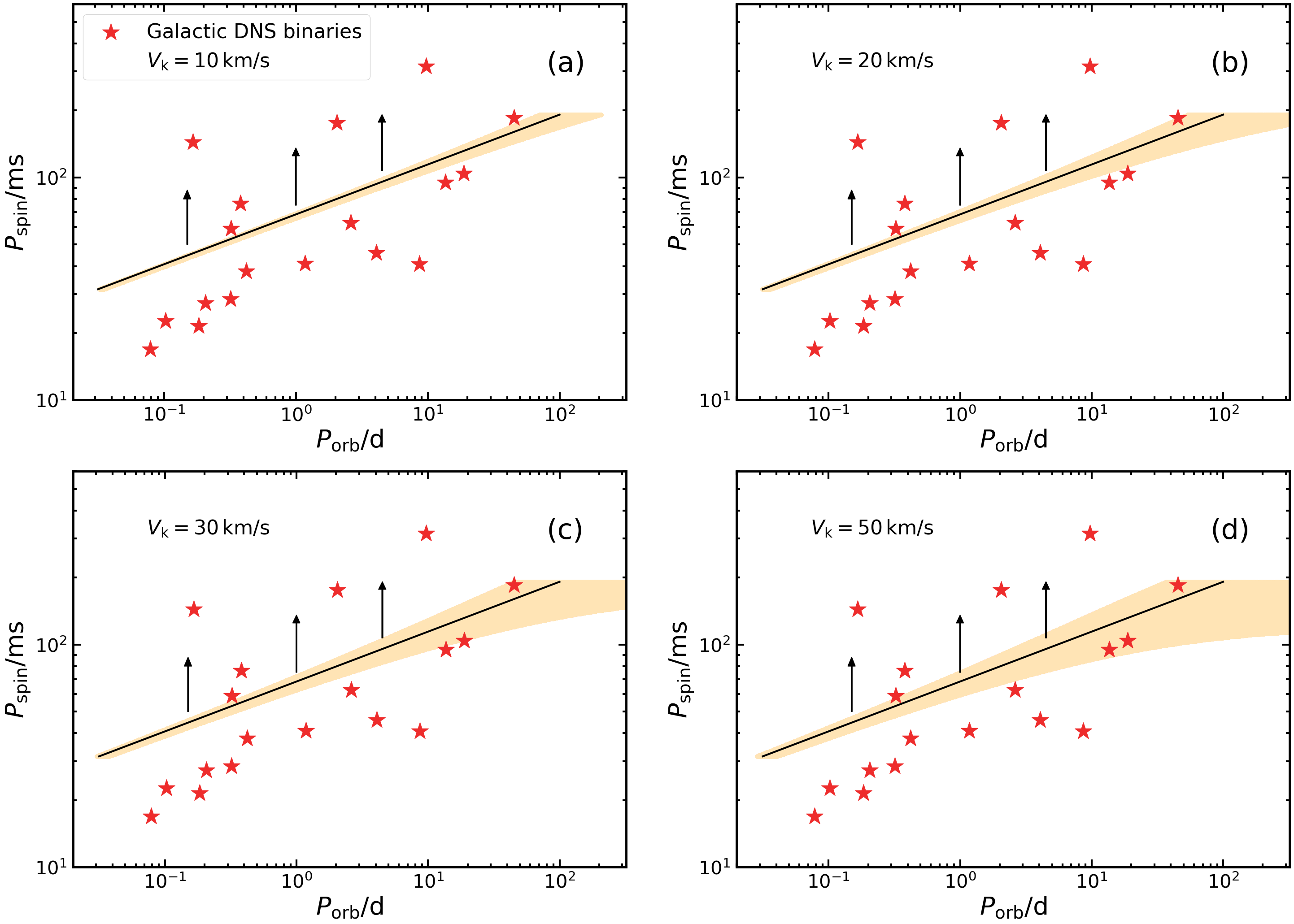}
	\caption{The $P_{\rm orb}-P_{\rm spin}$ diagram for the Galactic DNS systems (red stars) in observations.
		The black line (Equation \ref{con:ps-pb}) is a fit to our simulated results
for the ($P_{\rm spin}^{\rm min}, P_{\rm orb}^{\rm pre}$) correlation,
		and the brown shaded regions represent the effect of the NS kick velocities
		on the post-SN $P_{\rm orb}$.}
	\label{fig:pb-p0}
\end{figure*}

Fig.\,\ref{fig:pe} represents the distribution of the simulated post-SN systems
in the $P_{\rm orb}-e$ diagram, as well as the Galactic DNS systems in observations (red stars)\footnote{This probability distribution is not necessarily reflective of reality since we assume a uniform distribution of pre-SN log$P_{\rm orb}$.}.
We set $500$ pre-SN $P_{\rm orb}$ that vary logarithmically from $0.07$\,d to $50$\,d,
and each panel corresponds to different $V_{\rm k}$,
i.e. $10, 20, 30$ and $50\,\rm km\,s^{-1}$.
We note that the high probability features in this plot,
identified as lines I, II, and III,
trace the high density regions of similar identification in Fig.\,\ref{fig:kick-pe}
(corresponding to post-SN orbital configurations that are weakly dependent on the kick angles),
when convolved with variations in the orbital period and kick magnitude.
It can be seen that the NS kicks $\lesssim50\,\rm km\,s^{-1}$ could reproduce the most DNS systems,
especially for the binaries with relatively low eccentricities.
For example, PSR $\rm J1755-2550$ \citep[][]{2015MNRAS.450.2922N} is a DNS candidate with
a long orbital ($P_{\rm orb}=9.7$\,d) and a low eccentricity ($e=0.09$).
\citet{2018MNRAS.476.4315N} suggested that the minimum mass for
the companion star of PSR $\rm J1755-2550$ is $0.9\,\rm M_\odot$,
but it is not completely determined
whether the companion star is a NS or a WD.
By assuming that PSR $\rm J1755-2550$ is a DNS system,
\citet{2017ApJ...846..170T} suggested that
this system may be formed by a NS kick lower than $50\,\rm km\,s^{-1}$.
Our simulations also show that
a NS kick of $\lesssim30\,\rm km\,s^{-1}$ (most likely $\sim10\,\rm km\,s^{-1}$)
may explain the formation of PSR $\rm J1755-2550$.

PSR $\rm J1930-1852$ is the widest DNS system with a orbital period of $45$\,d
\citep[][]{2015ApJ...805..156S}.
From Fig.\,\ref{fig:pe}, it is seen that
its NS kick velocity was likely in the range of $\sim20-50\rm\,km\,s^{-1}$.
PSR J$1811-1736$ has a long orbital period ($P_{\rm orb}=18.8$\,d) and
the highest eccentricity ($e=0.83$) among the known Galactic DNS systems
\citep[][]{2007A&A...462..703C}.
Although the results show that PSR J$1811-1736$ can be produced by kicks of $\sim50\rm\,km\,s^{-1}$,
this source may come from different formation channel
because of the huge gap at long periods and moderate eccentricities.
In addition,
kicks lower than $\sim50\rm\,km\,s^{-1}$ cannot account for three DNSs with high eccentricity,
i.e. PSRs $\rm J0509+3801$, $\rm J1757-1854$ and $\rm J1915+1606$.
The distribution of the simulated post-SN systems for higher NS kick velocities are shown in Fig.\,\ref{fig:pe-fe}.
We can see that these three high eccentricity DNSs may originate from high NS kicks larger than $\sim150\,\rm km\,s^{-1}$
\citep[see also][]{2017ApJ...846..170T}.

\subsection{Properties of pre-SN systems}
Table \ref{table:2} shows the properties of Galactic DNS systems,
as well as their NS kick velocities and pre-SN orbital periods estimated in this work.
We found that
the DNSs with short orbital periods and high eccentricities
(e.g. PSRs J$0509+3801$, J$1757-1854$ and J$1915+1606$)
may be produced by high kick velocities
\citep[see also][]{2017ApJ...846..170T}.
However,
it has been suggested that such systems may originate from other formation channel
\citep[see][and discussion in Section 5.2]{2019ApJ...880L...8A}.
In addition,
half of the DNSs have pre-SN orbital periods less than $\sim1$\,d and
most DNSs may originate from the relatively low NS kicks.
This may be caused by the selection effects, that is,
the probability of surviving the pre-SN system is higher
if it has a short orbital period or a low kick velocity.
\citet{2017ApJ...846..170T} calculated the probability of surviving the second SN in a binary
with different pre-SN orbital periods and NS kicks.
Their results indicate that for relatively tight systems with $P_{\rm orb}^{\rm pre}\lesssim1\rm\,d$,
all systems will survive if the NS kicks are lower than $100\rm\,km\,s^{-1}$.
Meanwhile,
the orbital periods of such tight systems does not change significantly after the SN explosions,
which is helpful for studying their pre-SN properties.

\subsection{DNS systems in the $P_{\rm orb}-P_{\rm spin}$ diagram}
\subsubsection{The $P_{\rm orb}-P_{\rm spin}$ correlation}
NSs in close X-ray binaries could be spun up during the mass-accretion process
\citep[e.g.][]{1982Natur.300..728A, 1991PhR...203....1B},
and the minimum spin period of the recycled pulsars can be expressed as \citep[][]{2012MNRAS.425.1601T}\footnote{The initial spin angular momentum can be negligible owing to the spin-down process during the detached stage with a timescale of $\sim10^{6}$\,yr.
For example,
the NS with an initial spin period of $\sim10$\,ms
and a magnetic field of $10^{12}$\,G spins down on a timescale of
$\sim 10^{4}$\,yr due to magnetic dipole radiation
\citep[see also equation 2 in][]{2010ApJ...717..245K}.
In addition, the accretion will not play as significant a role
if pulsars are already spinning at a high rate
because pulsars have equilibrium periods.}:
\begin{equation}
	P_{\rm spin}^{\rm min} \approx 0.34\times(\Delta M_{\rm NS}/\rm M_\odot)^{-3/4}\,ms,
	\label{con:psmin-pb}
\end{equation}
in which $\Delta M_{\rm NS}$ is the accreted mass of NSs.
For Case BB RLOF,
the mass-transfer rate is much higher than $\dot M_{\rm Edd}$
(see Fig.\,\ref{fig:mass-loss}),
resulting in that most of the transferred material is ejected from the binaries
and only a small amount of material are accreted by NS.
From Table \ref{table:1},
we can see that the accreted masses of NSs range from
$2.0\times10^{-4}\,\rm M_\odot$ to $2.0\times10^{-3}\,\rm M_\odot$,
corresponding to initial orbital periods of $0.07-10$\,d.
This results in the formation of recycled pulsars
with $P_{\rm spin}^{\rm min}\sim36-210$\,ms after the He star companions undergo EC-SNe.
Table \ref{table:1} lists $P_{\rm spin}^{\rm min}$ and final orbital periods of NSs prior to EC-SNe,
and their relation can be written as:
\begin{equation}
	P_{\rm spin}^{\rm min} \approx 68\pm6\rm\,ms\times(\textit{P}_{\rm orb}^{\rm pre}\rm/d)^{0.22\pm0.05}.
	\label{con:ps-pb}
\end{equation}
In addition,
longer $P_{\rm orb}^{\rm i}$ and larger $M_{\rm He}^{\rm i}$
result in lower $\Delta M_{\rm NS}$,
thereby leading to longer spin periods of the recycled pulsars
\citep[see also][]{2015MNRAS.451.2123T}.

Fig.\,\ref{fig:pb-p0} represents the Galactic DNS systems (red stars) in the $P_{\rm orb}-P_{\rm spin}$ diagram.
The black line indicates the relation
between the simulated $P_{\rm spin}^{\rm min}$ and
$P_{\rm orb}^{\rm pre}$ prior to SN (see Equation \ref{con:ps-pb}).
From this figure,
we can see that the DNSs with longer $P_{\rm orb}$ are
preferentially accompanied by the recycled pulsars with higher $P_{\rm spin}$,
resulting from the shorter duration of the mass-accretion phase
\citepalias[e.g. \citealt{2015MNRAS.451.2123T};][]{2023MNRAS.526..932G}.
The brown shaded regions indicate the effect of NS kicks on the post-SN $P_{\rm orb}$.
As shown in this figure,
longer $P_{\rm orb}^{\rm pre}$ and higher NS kicks will result in a wider range of the post-SN orbital periods.
However,
we note that this results hardly explain the DNS systems
with relatively rapidly spinning recycled pulsars ($P_{\rm spin}\lesssim30$\,ms).
For example,
the DNS system PSR J$1946+2052$ has a spin period of $17$\,ms,
which is lower than the minimum value in our simulations
($\sim 38$\,ms, see Table \ref{table:1}).
It has been suggested that
such mildly recycled pulsars may be caused by the super-Eddington accretion,
that is,
Case BB RLOF allows first-born NS to accrete material
at mass-accretion rates of $2-3$ times $\dot M_{\rm Edd}$
\citep[e.g.][]{2014MNRAS.437.1485L, 2017ApJ...846..170T}.
In this work,
we also provide a possible explanation for the relatively low spin period of pulsars
by considering the residual H envelope on the surface of the He star companion, see Section \ref{h-en}.

\begin{table*}
		\centering
		\caption{
			Characteristics of the binaries considering the residual H envelope on the surface of the He star companion,
in which we set the stellar wind efficiency to be $0.1$.
$M_{\rm He}^{\rm i}$ is the mass of the He-ZAMS model with H envelope;
$P_{\rm orb}^{\rm i}$ is the orbital period after the artificial stripping event;
$M_{\rm He-core}$, $M_{\rm H-env}$ and log$R^{\rm i}$
are the initial He core mass, H envelope mass and the radius of the He star companion, respectively.
		}
		\label{table:3}
	\begin{tabular}{ c  c c c ccc c ccc c c c }
		\toprule
		\hline 
		Set		&&$M_{\rm He}^{\rm i}$  &log$P_{\rm orb}^{\rm i}$  &$M_{\rm He-core}$	&$M_{\rm H-env}$ & log$R^{\rm i}$&$M_{\rm core, f}$&$M_{\rm env, f}$& $\Delta M_{\rm NS}$ 	&$P_{\rm spin}^{\rm min}$ &$P_{\rm orb}^{\rm f}$  &$t_{\rm f}$	&Final fate\\
		&&($\rm M_\odot$)&   (d)& ($\rm M_\odot$)& ($\rm M_\odot$) &($\rm R_\odot$)& $(\rm M_\odot)$  &($\rm M_\odot$) &($\rm M_\odot$)& (ms) &(d)	&(Myr) &\\
		\hline 
		17	&&$2.66$		&$-1.00$	&$2.500$	&$0.160$&$-0.20$&$1.386$&1.2e-2&4.1e-3	&$21$	&$0.088$	&$2.25$	&EC-SN\\
        \hline 
		18	&&$2.73$		&$0$	&$2.128$	&$0.598$&$0.42$&$1.388$&1.0e-1&1.5e-3		&$45$	&$0.833$  &$2.66$	&EC-SN\\
		\hline 
		19	&&		$2.72$	&$1.00$	&$2.125$	&$0.595$&$0.36$&$1.470$&3.8e-1&1.0e-3
		&$60$	&$7.701$	&$2.46$		&Fe CC-SN\\
		\hline
20&&$2.73$&$0$&$2.73$&$0$ &$-0.37$&$1.513$&2.5e-1&3.6e-4 &$130$	&$0.780$&$2.30$		 &Fe CC-SN\\
\hline
	\end{tabular}
\end{table*} 

\subsubsection{The effect of residual H envelope on $P_{\rm spin}$}\label{h-en}
It is generally believed that
the NS+He star system that can form DNS has gone through the common envelope (CE) phase,
in which the dynamic friction generated by the motion of NS inside the giant star's  envelope leads to the loss of orbital angular momentum and the ejection of the H-rich envelope
\citep[e.g.][]{1976IAUS...73...75P, 2003MNRAS.344..629D, 2013A&ARv..21...59I, 2017ApJ...846..170T, 2023arXiv231107278W}.
However, recent simulations suggest that
a small amount of H shell still remains on the surface of He star companion
after CE ejection
\citep[e.g.][]{2019ApJ...883L..45F, 2022ApJ...933..137G, 2024ApJ...961..202G},
which may affect the evolution of NS+He star systems.
Accordingly,
we calculated the evolution of three binaries taking into account the residual H envelope.

We evolved a $11\rm\,M_\odot$ MS star to the stage
near the tip of red supergiant phase,
where central helium burning has not yet occurred.
We then removed the envelope until its radius less than the roche lobe radius,
e.g. for binary with $M_{\rm He}^{\rm i}\sim2.7\rm\,M_\odot$ and $P_{\rm orb}^{\rm i}=0.1$\,d,
the initial radius of companion star should be less than $\sim0.63R_\odot$.
Meanwhile,
the mass of the remaining H envelope should be in the range of $\sim0.1-0.6\rm\,M_\odot$
\citep[see, e.g.][]{2019ApJ...883L..45F},
consistent with the extended test of a $13\rm\,M_\odot$ primordial star based on
\citet{2022ApJ...933..137G}.
After that,
we put the He-ZAMS model with H envelope into the binary module
to calculate the binary evolution.
We built three He star models with different H envelope masses
just after the artificial stripping process,
in which the H abundance in the He core is assumed to be less than $0.01$.
Note that the outer layer is a mixture of H and He
rather than a pure H shell.
Table\,\ref{table:3} lists evolutionary properties of three NS+He star binaries
by considering the H envelope,
as well as a binary system with a naked He star companion for comparison.
We can see that the accreted mass of NS increases several times
if the residual H envelope is assumed to remain on the surface of the He star companion.
Meanwhile,
the residual H envelope can also affect the final fate of the He star companion.

\begin{figure}
	\centering\includegraphics[width=\columnwidth*3/3]{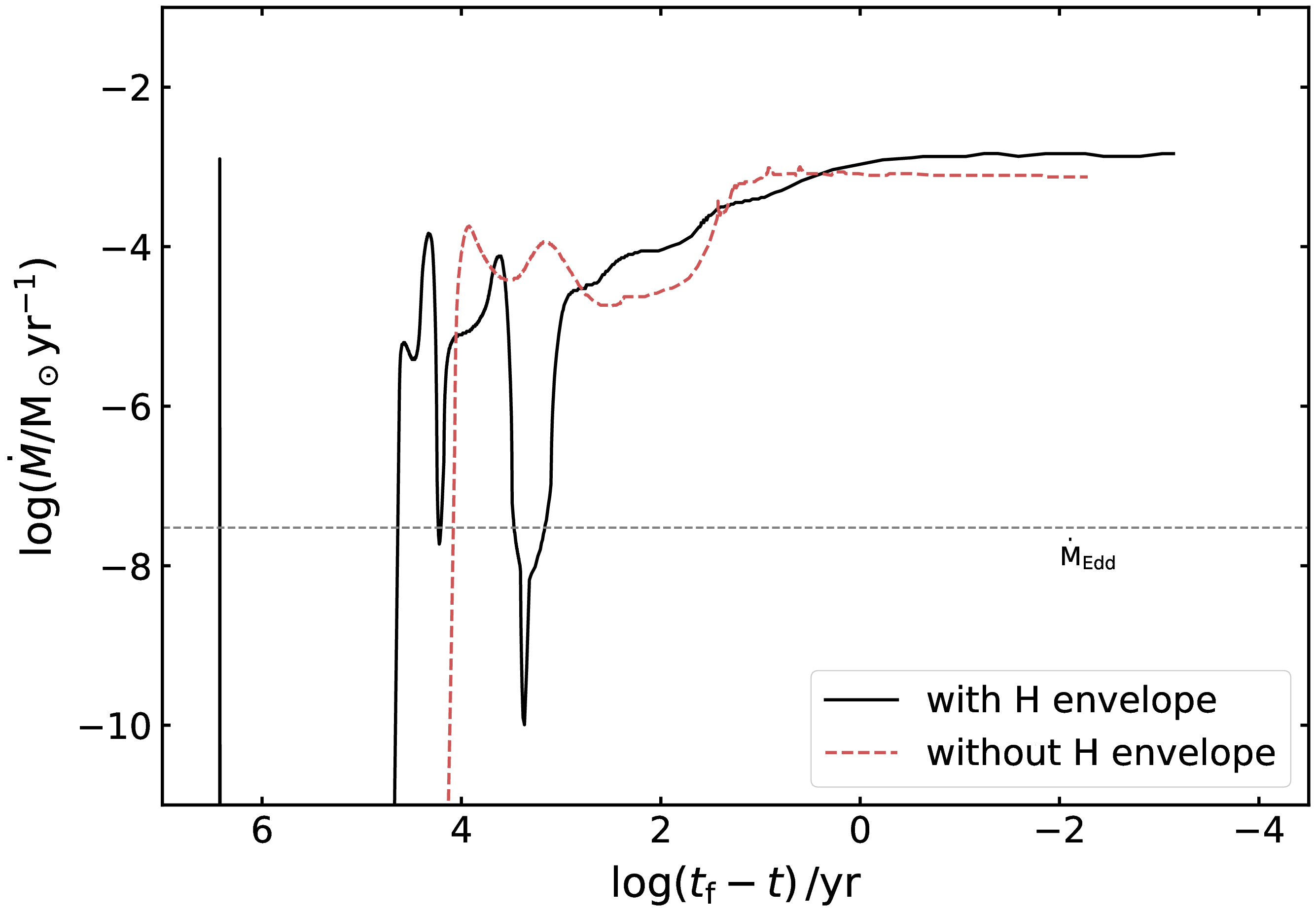}
	\caption{The effect of the H envelope on the mass-transfer rate, where the initial donor mass is $2.73\rm\,M_\odot$
(see sets 18 and 20 in Table\,\ref{table:3}).
The black line and the red-dashed line denote
the model considering the residual H envelope and the model with naked He star,
respectively.}
	\label{fig:h-enve-com}
\end{figure}

\begin{figure}
	\centering\includegraphics[width=\columnwidth*3/3]{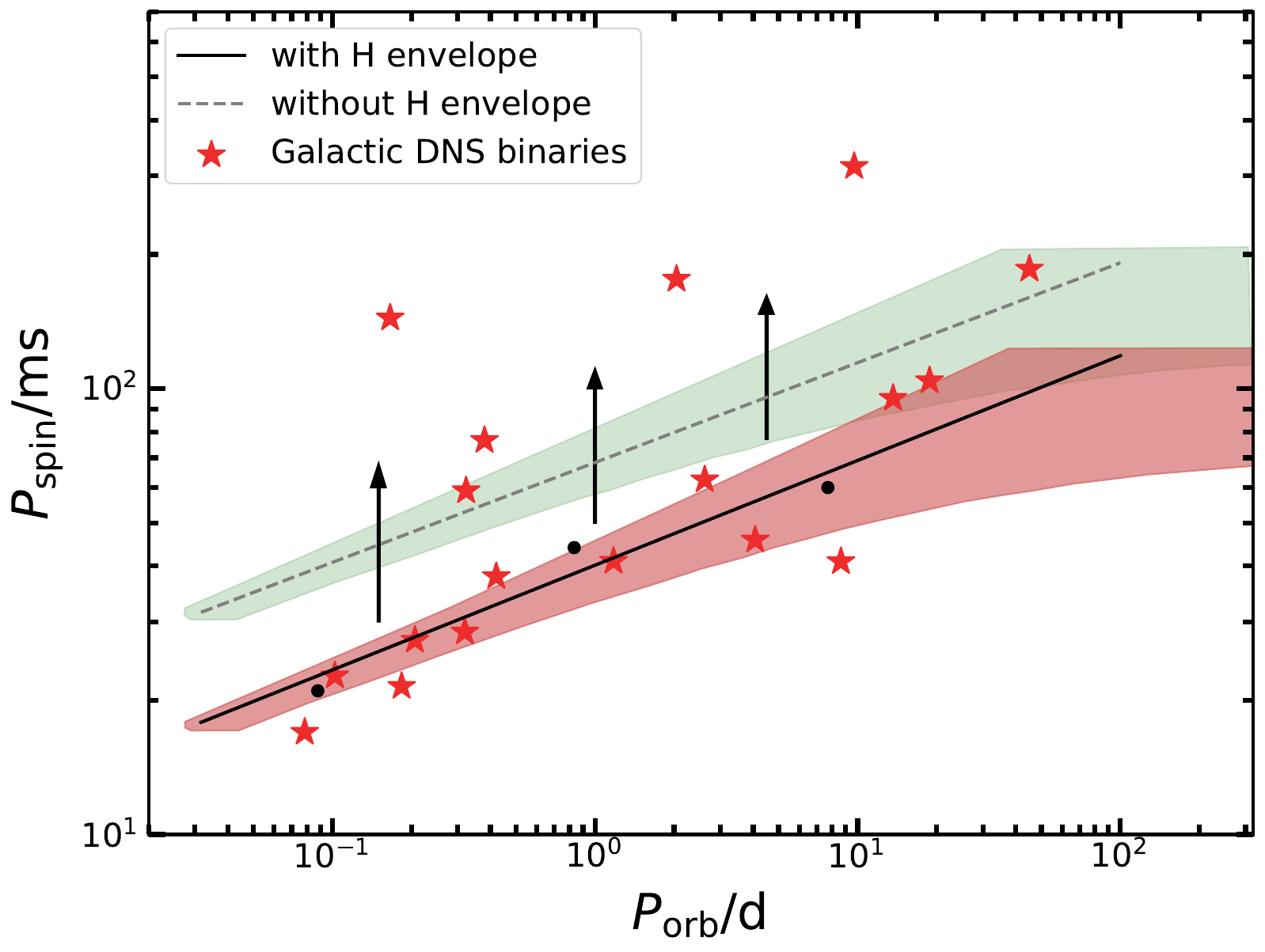}
	\caption{The $P_{\rm orb}-P_{\rm spin}$ diagram for the Galactic DNS systems (red stars) and our simulations.
The black line and the gray dashed line
represent the fitting results for the
models with and without residual H envelope, respectively.
The data of black dots come from sets $17-19$ in Table\,\ref{table:3}.
The green and red shaded regions represent the effect of the NS kick velocity
($V_{\rm k}=50\rm \,km\,s^{-1}$).
}
	\label{fig:h-en-pbp0}
\end{figure}
Fig.\,\ref{fig:h-enve-com} shows the effect of the residual H envelope on the mass-transfer rate,
in which the initial donor mass in both models is $2.73\rm\,M_\odot$
(see sets 18 and 20 in Table \ref{table:3}).
The He star with a H envelope has a larger radius than the naked He star,
causing it to fill its Roche lobe more easily.
It can be seen that
the H envelope results in a longer duration of the mass-transfer phase
\citep[][]{2021ApJ...922..245B},
thereby a larger accreted mass and a shorter spin period of NS.
Fig.\,\ref{fig:h-en-pbp0} shows the theoretical ($P_{\rm orb}$, $P_{\rm spin}$) correlation calculated with and without the residual H envelope.
The simulation results can better explain the observational characteristics of the DNS systems if we consider the H envelope on the surface of the He star companions.
It is worth noting that
the structure of the He star companion
and the mass of the H envelope depend on the process of CE ejection.
The impact of the residual H envelope on the evolution of pulsars
needs to be systematically studied in future.


\section{Discussion}
\subsection{Uncertainties in the modelling}
In our default models,
we adopted the `Dutch' stellar wind with a efficiency of $1.0$ to calculate the wind mass-loss rate.
However, different stellar wind efficiencies may change the evolutionary outcomes of the He stars.
To further explore the influence of stellar wind efficiencies on the evolution of NS+He star systems,
we computed four models with a stellar wind efficiency of $0.1$ (see Table\,\ref{table:1}).
The most significant effect is that lower wind efficiencies
result in lower initial mass of the He star companion for EC-SNe.
In addition,
we note that the remaining helium envelope mass ($M_{\rm He,f}^{\rm env}$)
increases with the decrease of wind efficiency
for the binaries with relatively long initial orbital periods,
e.g. $P_{\rm orb}^{\rm i}=10$\,d.
However,
the value of $M_{\rm He,f}^{\rm env}$ is still less than $0.06\,\rm M_\odot$
even though a low stellar wind efficiency is adopted,
indicating that the EC-SN is likely to be classified as a Type Ic SN
\citep[][]{2012MNRAS.422...70H}.
In addition,
convective overshooting is also an important physical process,
which can result in higher effective temperature/luminosity,
longer HeMS lifetime and a larger convective core
\citep[e.g.][]{2016RAA....16..141Y, 2022A&A...668A.106C}.
In the present work,
we set the overshooting parameter to be $0.014$
\citep[e.g.][]{2013ApJ...772..150J, 2022A&A...668A.106C}.
We expect that the initial parameter space for EC-SNe
will move up if a lower overshooting parameter is taken.

We assume that the ejected mass of the exploding star as $0.08\,\rm M_\odot$.
However,
by using the gravitational binding energy provided by \citet{1989ApJ...340..426L},
\citet{2015MNRAS.451.2123T} calculated the gravitational masses of the NS remnants,
and found that about $0.2\,\rm M_\odot$ of the material could be ejected
when the He star companions undergo EC-SNe.
Accordingly,
we calculated the distribution of post-SN systems in the $P_{\rm orb}-e$ diagram
by assuming the ejected mass as $0.2\,\rm M_\odot$ (see Fig.\,\ref{fig:pe-02}).
We found that
higher ejected mass leads to a wider range of eccentricity and the post-SN $P_{\rm orb}$,
although the influence is not significant.
In addition,
\citet{2017ApJ...846..170T} suggested that post-SN simulations
are also less dependent on the mass of the first-born NS.

\subsection{Comparison to previous studies}
\citet{2015MNRAS.451.2123T} studied the parameter spaces for
CO WDs, ONe WDs, EC-SNe and Fe CC-SNe in NS+He star systems.
They assumed that EC-SNe will be produced
if final metal core masses range from $1.37\,\rm M_\odot$ to $1.43\,\rm M_\odot$.
However,
our previous work show that the He star companions may explode as SNe Ia
if the metal core masses range from $\sim1.335-1.385\,\rm M_\odot$.
This could lead  to a $25\%$ reduction in the parameter space for producing EC-SNe,
which only applies for interacting He stars in close binaries
\citepalias[][]{2023MNRAS.526..932G}.
\citet{2018MNRAS.481.1908K} studied
the formation rate of DNSs by performing the binary population syntheses.
Their simulation results show that
more systems can survive even if they experience Fe CC-SNe with larger kicks,
given that most systems are tightened by CE process prior to the second SNe.
For the second SNe,
the ratio of EC-SNe to Fe CC-SNe is $\sim0.2$ in all DNSs formed.
Thus, the reduced EC-SN parameter space has no
significant impact on the formation rate of DNSs
owing to the much smaller contribution of EC-SNe to the second-born NSs.
In addition,
by comparing with the results of \citet{2015MNRAS.451.2123T},
we note that the remaining helium envelope mass calculated in their study
is larger than that of this work,
and their simulations show that
EC-SNe in some NS binaries may appear as ultra-stripped Type Ib SNe.
This is because \citet{2015MNRAS.451.2123T}
stopped the code when the ONe cores are formed,
while we continued the calculation to a later evolution stage,
i.e. the $e$-capture on $\rm ^{20}Ne$ (see Fig.\,\ref{fig:ele-ob}).
We note that the donors still lose material
at a rate of $\sim10^{-4}-10^{-3}\rm\,M_\odot\,\rm yr^{-1}$
from ONe core formation to the $e$-capture on $\rm ^{20}Ne$
(see Fig.\,\ref{fig:mass-loss}),
which reduces the donor mass by $\sim0.1\rm\,M_\odot$.
Our simulations show that the final He envelope masses are lower than $0.06\rm\,M_\odot$ for NS+He star binaries with $P_{\rm orb}^{\rm i} \lesssim 10$\,d,
and such EC-SNe may be observed as Type Ic SNe.

\citet{2018ApJ...867..124S} investigated the effect of the NS kicks
on the distribution of DNS systems with population synthesis studies.
They suggested that the second-born NSs in most DNS systems
have kick velocities less than $\rm 80\,km\,s^{-1}$.
\citet{2017ApJ...846..170T} explored the pre-SN characters of DNS systems,
and obtained the distribution of parameters of each source.
Their results indicate that the kick velocity and pre-SN orbital period
cover a wide range,
and the peak values of kick velocities for most of the observed DNS systems tend to be less than $\sim 50\rm\,km\,s^{-1}$
(see their Figs $24-38$).
Previous studies show that EC-SNe have explosion energies about $\sim10^{50}$\,erg and low ejected mass between $0.01-0.2\rm\,M_\odot$
\citep[e.g.][]{2006A&A...450..345K, 2006ApJ...644.1063D, 2022MNRAS.513.1317Z},
which may lead to the small NS kicks $\lesssim50\rm \,km\,s^{-1}$
or even $\lesssim10\rm \,km\,s^{-1}$
\citep[e.g.][]{2017ApJ...837...84J, 2017ApJ...846..170T}.
Meanwhile,
our results show that the NS kick velocities of
$\lesssim50\rm\,km\,s^{-1}$
can produce most observed DNS systems,
indicating that EC-SNe may play an important role in the formation of DNSs.

\citet{2017ApJ...837...84J} studied how the NS kick velocity depends on
the explosion energy, ejecta mass, NS mass,
and other relevant parameters.
They found that massive metal cores tend to result in larger ejected masses
and higher explosion energies, and therefore higher NS kick velocities.
Meanwhile,
\citet{2017ApJ...846..170T} made statistics related to
the masses of the second-born NSs and the kick velocities.
They found that the second-born NSs with masses $\gtrsim 1.33\,\rm M_\odot$
are more likely to have high kick velocities,
and suggested that smaller NS masses are preferentially accompanied by smaller NS kicks.
Table\,\ref{table:2} lists the NS kick velocities estimated from this work
and the masses of the second-born NSs in observations.
We also note that the DNS systems with high second-born NS masses
may originate from high NS kicks
(e.g. PSRs J$0509+3801$ and J$1757-1854$).

	\citet{2019ApJ...880L...8A} divided the observed DNSs into three distinct subtypes
	according to the orbital characteristics:
	(1) DNSs with tight orbits and low eccentricities,
	which can merge within the Hubble time;
	(2) DNSs with wide orbits that cannot merge within the Hubble time;
	(3) DNSs with tight orbits and high eccentricities.
	Our theoretical results show that
	both subtypes (1) and (2) can be reproduced by
	the NS kicks $\lesssim50\,\rm km\,s^{-1}$.
	To explain the clustering of DNSs in subtype (3),
	i.e. PSRs $\rm J0509+3801$, $\rm J1757-1854$ and $\rm J1915+1606$,
	\citet{2019ApJ...880L...8A} suggested that
	if the DNSs are the product of isolated binary evolution,
	then the second-born NSs must have a small natal kicks $\lesssim25\,\rm km\,s^{-1}$
	and the pre-SN He star masses narrowly distributed around $3.2\,\rm M_\odot$.
	In addition,
	they also proposed that the DNSs in subtype (3) may originate from dynamical formation channel,
	that is,
	the DNSs firstly formed in globular cluster,
	and was then dynamically kicked into the Milky Way field
	\citep[see][]{1991Natur.349..220P}.


PSR J$1952+2630$ is a mildly recycled $20.7$\,ms pulsar
orbiting together with a massive WD companion ($\gtrsim0.93\rm\,M_\odot$),
and its orbital period is $9.4$\,h
\citep[][]{2014MNRAS.437.1485L}.
Such tight orbit indicates that this system has undergone CE evolution and form a NS+He star system,
and then the NS spins up through the Case BB RLOF.
\citet[][]{2014MNRAS.437.1485L} suggested that
the relatively fast spin of this pulsar may be caused by the super-Eddington accretion
with mass-accretion rate between $\sim100-300$ per cent of $\dot M_{\rm Edd}$.
According to our simulations,
we concluded that
such mildly recycled pulsar+WD systems can also be produced
if the residual H envelope on the surface of He star companion is considered.

\subsection{Thermonuclear EC-SNe}
Previous studies usually assumed that
the $e$-capture on $\rm ^{20}Ne$ takes place
if ONe cores have masses $\gtrsim1.37\,\rm M_\odot$,
resulting in that the ONe cores collapse into NS through EC-SNe
\citep[e.g.][]{1984ApJ...277..791N,2015MNRAS.451.2123T,2015MNRAS.446.2599D}.
However,
by performing the multidimensional hydrodynamic simulations for oxygen deflagration
in ONe cores with three different central ignition densities,
\citet{Jones2016} found that
the ONe cores may not collapse into NSs
if oxygen deflagration is triggered
at ignition density $\rm log_{10}(\rho_{\rm c}/g\,cm^{-3})\lesssim10$,
while $\sim 0.1-1\rm\,M_\odot$ of material will be ejected
and leave bound ONeFe WD remnants,
called the thermonuclear EC-SNe
\citep[see also][]{1991ApJ...372L..83I, 2019PhRvL.123z2701K, Jones2019A&A}.
On the other hand, the ONe cores will collapse into NSs
at relatively high ignition density $\rm log_{10}(\rho_{\rm c}/g\,cm^{-3})\gtrsim10$.
Fig.\,\ref{fig:pc-tc1} shows that
the explosive oxygen burning occurs at central density of $\rm log_{10}(\rho_{\rm c}/g\,cm^{-3})\sim9.8$,
indicating that this NS+He star binaries may evolve into
eccentric pulsar+WD systems after SN explosions.

PSR $\rm J0453+1559$ is a DNS system candidate
with $P_{\rm orb} = 4.07$\,d and $e = 0.1125$ \citep{2015ApJ...812..143M}.
The mass of the recycled pulsar and its companion are
$1.559(5)\,\rm M_\odot$ and $1.174(4)\,\rm M_\odot$, respectively.
However, it is unclear whether the companion is a NS or a WD.
The mass of its companion is significantly lower
than that of the second-born NSs in other DNS systems,
indicating that the companion is more likely not a NS.
On the other hand,
the eccentricity of $0.1125$ is high for normal NS+WD systems.
In view of this, \citet{2019ApJ...886L..20T} used the thermonuclear EC-SN channel
to explain the formation of this system,
and they found that PSR $\rm J0453+1559$ may be produced
by a kick velocity from $\sim70\rm\,km\,s^{-1}$ to $100\rm\,km\,s^{-1}$.
More numerical simulations and observations about
thermonuclear EC-SNe are needed to verify this scenario.
\section{Summary}
Using the stellar evolution code MESA,
we systematically studied EC-SNe in NS+He star binaries with $P^{\rm i}_{\rm orb}<10$\,d.
In our simulations, the $e$-captures on $\rm^{20}Ne$ are triggered if the He star companions
develop highly degenerate ONe cores with masses from $\sim1.385\rm\,M_\odot$ to $1.43\rm\,M_\odot$.
SNe Ia are produced if the core masses $\lesssim1.385\rm\,M_\odot$,
owing to the explosive oxygen burning
caused by the convective Urca process.
The final fates of the He stars will be Fe CC-SNe if the core masses $\gtrsim1.43\rm\,M_\odot$.
We calculated a series of NS+He star binaries and obtained initial parameter spaces for producing EC-SNe
in the log$P_{\rm orb}^{\rm i}-M_{\rm He}^{\rm i}$ diagram,
and found that higher metallicity can result in
higher $M_{\rm He}^{\rm i}$ and minimum $P_{\rm orb}^{\rm i}$ for EC-SNe.

By considering the kicks added to the second-born NS,
we then explored the properties of the formed DNS systems
after the He star companions collapse into NSs through EC-SNe,
such as the post-SN orbital period, eccentricity
and the spin period of recycle pulsar, etc.
Based on the comparison between numerical simulations and observations,
we found that the majority of the observed DNS systems can be explained by the NS kicks of $\lesssim50\rm\,km\,s^{-1}$,
indicating that the EC-SN channel may play an important role in the formation of DNS systems.
We also estimated the properties of the pre-SN systems of the observed DNS systems,
and found that half of the observed DNS systems seem to have tight pre-SN orbit ($\lesssim1$\,d),
resulting from the higher probability of surviving post-SN systems.
In addition,
by assuming that the residual H envelope remains on the surface of the He star companion,
we found that the first-born NSs could accrete enough material to form
the mildly recycled pulsars ($P_{\rm spin}\sim20$\,ms) in DNS systems.

\section*{Acknowledgements}
We thank the anonymous referee for valuable comments
that help to improve the paper.
This study is supported by the
the National Natural Science Foundation of China (Nos 12041301, 12121003, 12225304, 12090040/12090043, 12273014, 12288102, 12125303, 12173081 and 11733009),
National Key R\&D Program of China (Nos 2021YFA0718500, 2021YFA1600404 and 2021YFA1600403),
the Western Light Project of CAS (No. XBZG-ZDSYS-202117),
the science research grant from the China Manned Space Project (No. CMS-CSST-2021-A12),
the Yunnan Fundamental Research Project (Nos 202101AV070001 and 202201BC070003),
the Frontier Scientific Research Program of Deep Space Exploration Laboratory (No. 2022-QYKYJH-ZYTS-016),
the key research program of frontier sciences, CAS, No. ZDBS-LY-7005,
the International Centre of Supernovae, Yunnan Key Laboratory (No. 202302AN360001),
and the Shandong Fundamental Research Project (No. ZR2021MA013).
The authors also acknowledge the ``PHOENIX Supercomputing Platform''
jointly operated by the Binary Population Synthesis Group
and the Stellar Astrophysics Group at Yunnan Observatories, Chinese Academy of Sciences.
\section*{Data availability}
Results will be shared on reasonable request to corresponding author.

\bibliographystyle{mnras} 
\bibliography{1bib.bib}					


\newpage
\appendix
\section{The effect of higher kicks and ejected mass}
\begin{figure*}
	\centering\includegraphics[width=\columnwidth*9/5]{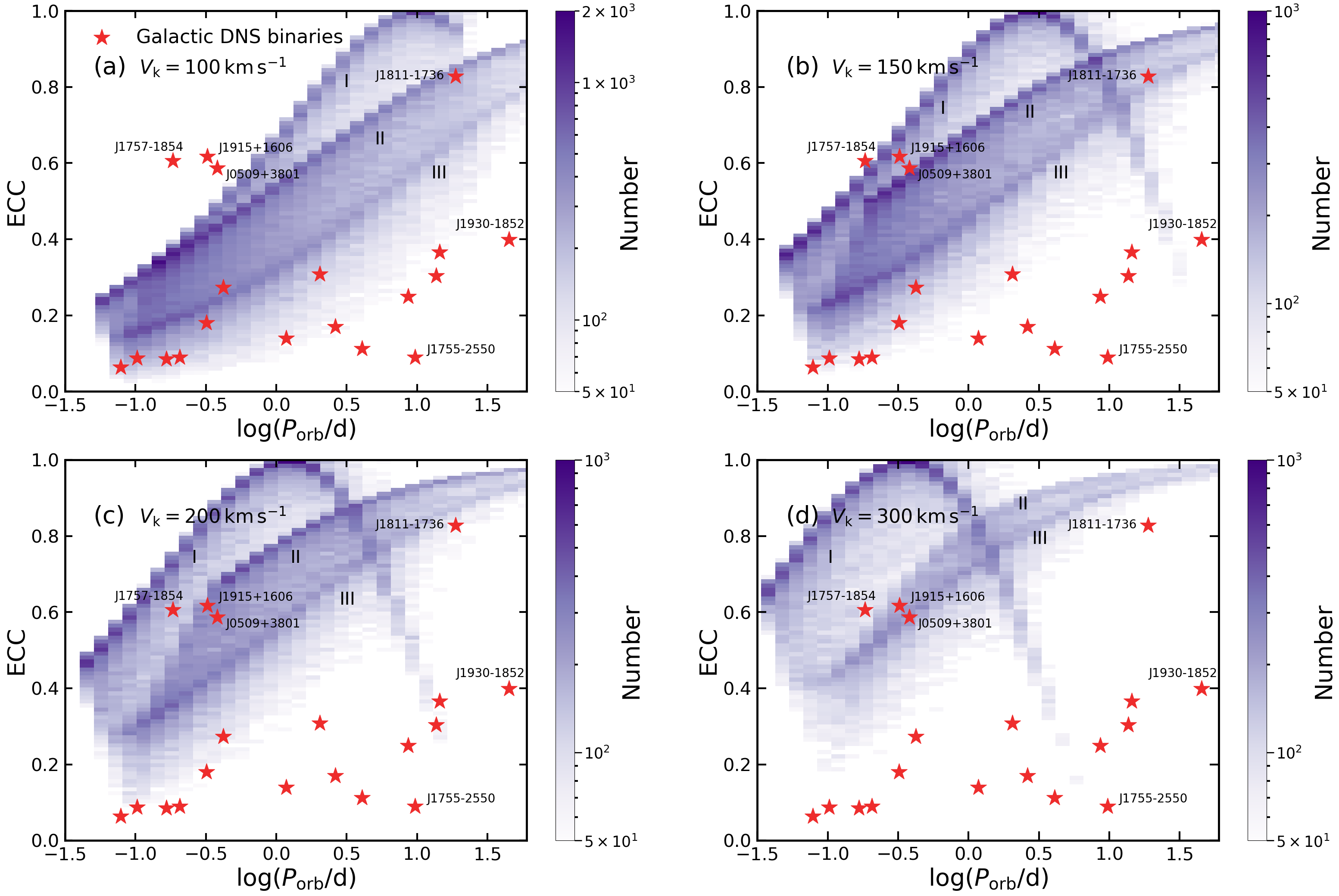}
	\caption{Similar to Fig.\,\ref{fig:pe}, but for high NS kick velocities,
		i.e. $V_{\rm k}=100, 200, 300$ and $\rm 400\,km\,s^{-1}$.}
	\label{fig:pe-fe}
\end{figure*}
\begin{figure*}
	\centering\includegraphics[width=\columnwidth*9/5]{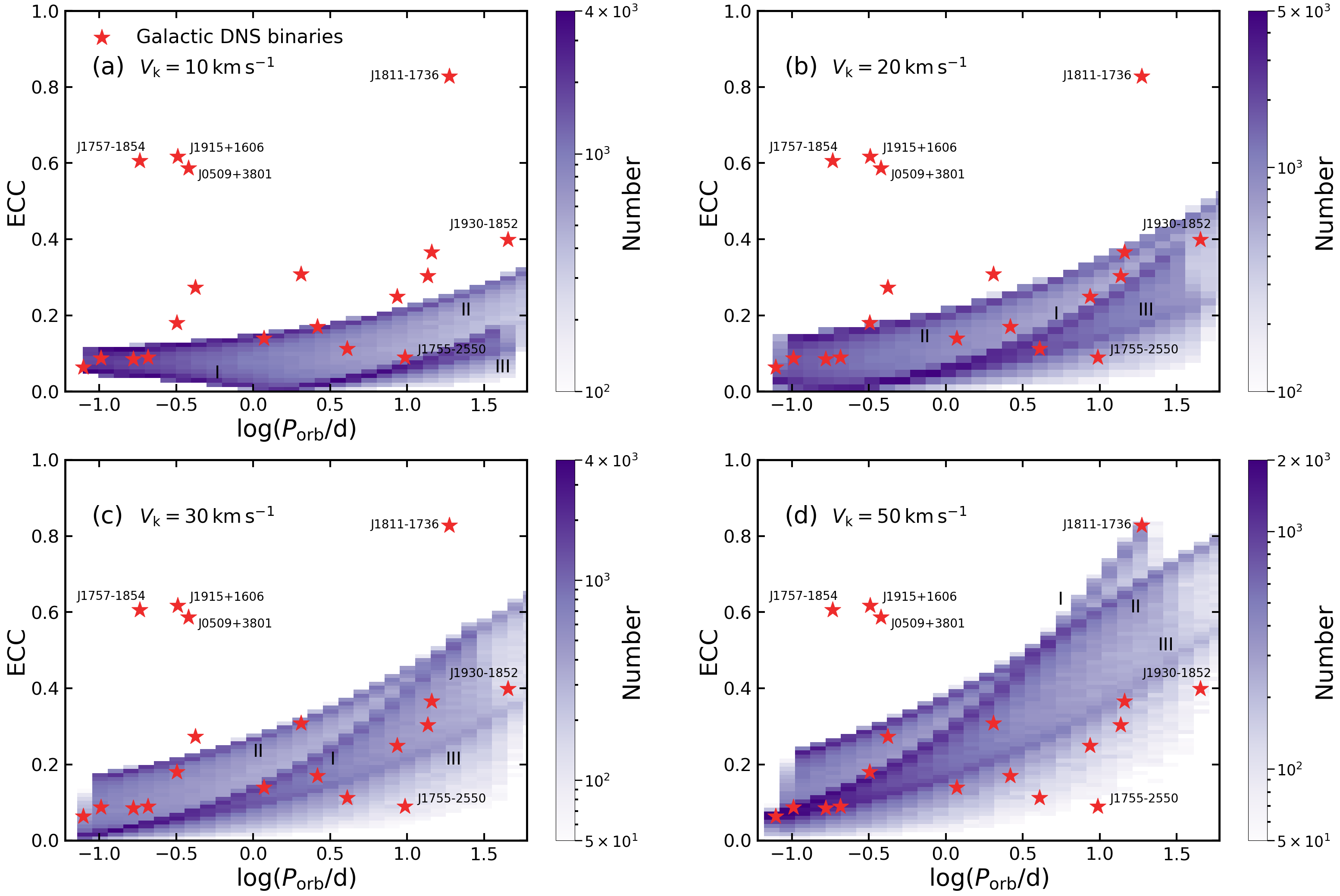}
	\caption{Similar to Fig.\,\ref{fig:pe}, but for $\Delta M=0.2\rm\,M_\odot$.}
	\label{fig:pe-02}
\end{figure*}

\label{lastpage}
\end{document}